\documentclass[twocolumn,trackchanges]{aastex63}

\received{}
\revised{}
\accepted{}
\submitjournal{ApJ}

\shorttitle{Bunch Particle Interaction in Pulsar Magnetospheres}
\shortauthors{Ben\'{a}\v{c}ek et al.}
\begin{document}

\title{Bunch Expansion as a cause for Pulsar Radio Emissions}

\correspondingauthor{Jan Ben\'a\v{c}ek}
\email{benacek@tu-berlin.de}

\author[0000-0002-4319-8083]{Jan Ben\'a\v{c}ek}
\affiliation{Center for Astronomy and Astrophysics, Technical University of Berlin, 10623 Berlin,
Germany}
%\nocollaboration{1}

\author[0000-0002-3678-8173]{Patricio~A.~Mu\~noz}
\affiliation{Center for Astronomy and Astrophysics, Technical University of Berlin, 10623 Berlin,
Germany}
%\nocollaboration{1}

\author[0000-0002-5700-987X]{J\"org~B\"uchner}
\affiliation{Center for Astronomy and Astrophysics, Technical University of Berlin, 10623 Berlin,
Germany}
\affiliation{Max Planck Institute for Solar System Research, 37077 G\"ottingen, Germany}



\begin{abstract}
Electromagnetic waves due to electron-positron clouds (bunches), created by cascading 
processes in pulsar magnetospheres, have been proposed to explain the pulsar radio 
emission.
In order to verify this hypothesis, we utilized for the first time Particle-in-Cell (PIC-) code 
simulations to study the nonlinear evolution of electron-positron bunches in dependence
on the relative drift speeds of electrons and positrons, on the initial plasma temperature, 
and on the initial distance between the bunches.
For this sake, we utilized the PIC-code ACRONYM with a high-order field solver and particle 
weighting factor, appropriate to describe relativistic pair plasmas. 
We found that the bunch expansion is mainly determined by the relative electron-positron
drift speed.
Finite drift speeds were found to cause the generation of strong electric fields that reach  up
to $E \sim 7.5 \times 10^{5}$~V/cm ($E / (m_\mathrm{e} c \omega_\mathrm{p} e^{-1}) \sim 4.4$)
and strong plasma heating.
As a result, up to 15~\% of the initial kinetic energy is transformed into the electric field energy. 
Assuming the same electron- and positron-distributions we found that the fastest 
(in the bunch reference frame) particles of consecutively emitted bunches eventually 
overlap in the momentum (velocity) space.
This overlap causes two-stream instabilities that generate (electrostatic) subluminal 
L-mode waves with electric field amplitudes reaching up to 
$E \sim 1.9\times 10^{4}$~V/cm ($E / (m_\mathrm{e} c \omega_\mathrm{p} e^{-1}) \sim 0.11$).
We found that the interaction of electron-position bunches 
leads to plasma heating, to the generation of strong
electric fields and of intense superluminal L-mode waves
which, in principle, can be behind the observed 
electromagnetic emissions of pulsars in the radio wave 
range.
\end{abstract}

\keywords{}


%%%%%%%%%%%%%%%%%%%%%%%%%%%%%%%%%%%%%%%%%%%%%%%%%%%%%%%%%%
%%%%%%%%%%%%%%%%%%%%%%%%%%%%%%%%%%%%%%%%%%%%%%%%%%%%%%%%%%

\section{Introduction} 
\label{sec:intro}

Pulsars are strongly magnetized  ($10^{8} - 10^{13}$~G), 
fast rotating neutron stars that emit radio waves into narrow cones.
Their emission is considered to be generated in the relativistically hot electron-positron 
magnetospheric plasma, studied for more than fifty years now~\citep[see, e.g.,][]{Kramer2002,Beskin2018}.
A number of models has been proposed to explain the observed radio waves
and their properties~\citep{Sturrock1971,Ruderman1975,Usov1987,Petrova2009,Philippov2020}.
Nevertheless, no consensus has been reached about the proper emission mechanism, the
efficiency of the energy conversion, and its relation to the observed pulsar radio 
signals~\citep{Melrose2017a,Melrose2020a}.

Current ``standard'' models of the pulsar magnetospheric plasma are based on the 
formation of Goldreich-Julian currents \citep{Goldreich1969,Cheng1977a}
that screen the convective electric field in closed magnetic-field-line regions in
a way that $\mathbf{E}\cdot\mathbf{B} = 0$ holds. On the other hand, over the polar 
caps with open magnetic field lines, where particles can escape the pulsar 
magnetospheres, electric field components parallel to the magnetic field can
be formed. In these regions known as ``gaps'', strong parallel electric field 
components ($\mathbf{E}\cdot\mathbf{B} \neq 0$) with $E  \approx 10^{12}$~V/m 
can exist.

If the plasma density is small, particles
in the gap region are accelerated to ultrarelativistic velocities with Lorentz 
factors $\gamma \sim 10^{6}-10^{8}$, forming the so-called ``primary beams'' 
which emit $\gamma$-ray photons.
The $\gamma$-ray  photons propagate into arbitrary directions,
eventually interacting with the magnetic field.
As a result they decay into electron-positron 
pairs:
$\gamma\mathrm{-photon} + \mathbf{B} \rightarrow e^+ + e^-$ \citep{Cheng1977a,Buschauer1977}.
The newly formed ``secondary beams'' of electrons and
positrons are denser than the primary beams by a 
factor $10^2 - 10^4$
and their Lorentz factor is rather 
$\gamma \sim 10^2 - 10^3$ \citep{Arendt2002}.
Initially their velocity has parallel and perpendicular 
to the magnetic field components.
The perpendicular velocity component quickly vanishes;
however, by emitting synchrotron radiation within
short times ($\ll \omega_\mathrm{p}^{-1}$, 
where $\omega_\mathrm{p}$ is the plasma 
frequency).
Hence, only their magnetic-field aligned velocity component 
remains.
The release of secondary particles is not a continuous process.
As the secondary beam particles escape their generation region at
relativistic speeds, the corresponding currents do not any more 
screen the electric fields.
Hence, their sparking repeats again and again at time scales of 
$\sim 500$~ns \citep{Ruderman1975,Cheng1977b}.

The interaction of the bunches of electrons and positrons 
(sometimes called clouds) can generate plasma waves due
to streaming instabilities which might emit radio 
waves~\citep{Eilek2016,Mitra2017,Melrose2017b,Melrose2020a}.
Three basic scenarios (C1-C3) of the interaction between the
particle bunches of primary and secondary beams  
have been suggested so far:

\begin{enumerate}
 \item[C1:] Relative streaming of the primary and secondary beam particles.
 A reactive instability with large growth rates could develop in the lower pulsar 
 magnetosphere within a distance $10-100 R_\star$, where $R_\star$ 
 is the neutron star radius \citep{Cheng1977b,Buschauer1977}.
 However, this instability has been found to be inefficient due to its small
 growth rates \citep{Benford1977,Arons1981} due to relativistic effects.
 
 \item[C2:] Acceleration of the secondary particles due to the net current.
 Electrons and positrons are accelerated in opposite directions, thus 
 possibly causing a two-stream instability \citep{Cheng1977a}. 
 Thermal effects were taken into account by \citep{Weatherall1994,Asseo1998}.
 The instability calculations assumed an infinitely long interaction
 region with a constant particle density. When taking the spatial size of the 
 bunch into account, the drift velocity between species separates the location 
 of oppositely charged particles and creates electric discharges in the course
 of their evolution. 
 Thus the constant density approximation can be only sufficient for a large bunch or very short times $<\omega_\mathrm{p}^{-1}$.
 
 \item[C3:] Relative streaming of the bunch particles.
The sparking processes can create ``trains'' of hot bunches that expand 
in the course of their propagation along the pulsar magnetic fields.
 As a result a streaming instability can occur when the fastest particles 
of a trailing bunch catch up with the particles of the leading 
bunch~\citep{Usov1987,Ursov1988,Asseo1998}.
Note that a typical bunch size is $L \approx 30l \approx 0.3 R_\star \approx 3$~km, 
where $l$ is the distance between bunches~\citep{Usov2002}.
\citet[Appendix A]{Melrose2020a} pointed out that, although the time needed for 
catching up slow particles of the leading bunch is sufficiently short,  the time for catching 
up with particles moving at the mean velocity is too long for a viable pulsar emission mechanism.
\end{enumerate}

A production of secondary electron and positron beams is necessary for 
scenarios C2 and C3 to work. It has been studied by several authors.
\citet{Arendt2002}, e.g., derived the properties of the produced secondary pairs.
According to them, the beam Lorentz factors range from $10^2$ to $10^3$ and 
the plasma inverse temperature is $\rho = m_\mathrm{e}c^2 / k_\mathrm{B}T=1.0-1.5$, depending
on the photon energy of pulsars, assuming a magnetic field of the neutron star equal to $B_\star = 10^{12}$~G.
Though the estimated inverse temperature is in a quite narrow range,
e.g. \citet{Weatherall1994,Rafat2019c} consider higher pulsar inverse 
temperatures, reaching up to $\rho \approx 0.01$.
\citet{Philippov2020} numerically simulated the formation of electron-positron pair 
bunches by a cascading process. The cascading can occur along the open magnetic 
field at heights where the magnetic field is small enough to allow a
velocity component perpendicular to the magnetic field direction, which 
is not immediately radiated.
\citet{Cruz2020} investigated the development of the pair cascade analytically.
They confirmed their analytical results utilizing one-dimensional PIC-code simulations 
taking quantum electrodynamic effects into account 
Their studies revealed the formation of particle clouds associated with large 
amplitude electrostatic waves.

\citet{Rafat2019c,Rafat2019a,Rafat2019b} studied  dispersion properties of 
a relativistically hot pair plasma and the consequent streaming instabilities 
in a pulsar magnetosphere. They found that the waves in a relativistic hot
pulsar magnetosphere manifest quite different dispersion properties compared 
to their non-relativistic counterparts. 
In particular that the generated L-mode  waves have superluminal and 
subluminal dispersion branches.
Most unstable is the subluminal L-mode wave close to the light line.
These authors calculated the corresponding growth rates of the different modes
associated to the streaming instabilities
and concluded that they are too small to generate sufficiently strong emission
in the lower pulsar magnetosphere.

\cite{Rahaman2020} compared the growth rates of the linear instabilities obtained for 
scenarios C2 and C3. 
They found that the growth rates of C2 can largely exceed those obtained of C3 
for certain multipolar magnetic field configurations at a
height of a few $R_\star$ above the pulsar surface.
For C3, however, the growth rates can be sufficiently large only for
$\kappa_\mathrm{GJ} \geq 10^4$, where $\kappa_\mathrm{GJ}$ is 
the multiplicity factor.

\citet{Manthei2021} obtained the properties of the relativistic streaming instabilities
for the scenario C3 by solving the linear dispersion relations 
and verifying them by PIC-code simulations.
They could confirm their linear-dispersion-theory predictions by their numerical
plasma simulations during the initial phase of the instability development, until
nonlinear effects prevail.
And also that sufficiently large growth rates
can be obtained for certain pulsar plasma parameters.
In particular they showed that an instability can develop at time-scales 
$<10\,\mu$s for $13<\gamma < 300$, $\rho_0;\rho_1 \geq 1$, 
and $n_1 / (\gamma_\mathrm{b} n_0) \geq 10^{-3}$, 
where $\rho_0,\rho_1$ are the inverse temperatures and 
$n_1, n_0$ are the densities of the background and the beam, 
respectively. $\gamma_\mathrm{b}$ is the beam Lorentz factor.

\citet{Benacek2021} studied the nonlinear stage of the relativistic pair-plasma 
streaming instability by means of PIC-code simulations.
They found that nonlinear superluminal L-mode waves can be formed  in
sufficiently cool plasmas $\rho \geq 1.66$, or for sufficiently large beam 
velocities $\gamma > 40$.
These waves are generated by a nonlinear interaction between the initially 
generated subluminal L-mode waves and an accompanying density wave.
These waves stay stable for sufficiently long times so that they can cause coherent 
electromagnetic (radio-) wave emissions.
These authors estimated the electrostatic energy density, which can be reached this way,
to be $1.2\times 10^{4}$~erg/cm$^3$ for subluminal waves and 
up to $1.1\times 10^{5}$~erg/cm$^3$ for superluminal waves.

As far as the mechanisms of radio wave emission are concerned, three
suggestions were made in the framework of scenarios C2 and C3~\citep{Melrose1995,Eilek2016,Melrose2017a,Melrose2020a}:

\begin{itemize}
 \item Relativistic plasma emissions can be caused by a modulational 
 instability of longitudinal electrostatic L-mode waves, which propagate 
 parallel to the magnetic field generating escaping electromagnetic waves 
 via a wave-wave interaction
 $T$, $L + M \rightarrow T$ \citep{Weatherall1997,Weatherall1998}.
 The modulational wave $M$ has a polarization vector component
 perpendicular to the magnetic field; the O-mode waves can escape the emission region.
 
 \item Linear acceleration and free electron maser/laser emission 
 mechanisms based on charges that undergo acceleration in the parallel or 
 in an arbitrary direction to the magnetic field lines, respectively.
A simplified model assumes particles oscillating with frequency 
$\omega_0$ and an electromagnetic emission at the frequency 
$\gamma_\mathrm{s}^2 \omega_0$, where $\gamma_\mathrm{s}$ is 
the particle Lorentz factor in the pulsar reference 
frame~\citep{Cocke1973,Melrose1978,Kroll1979,Fung2004,Melrose2009a,Melrose2009b,Reville2010,Timokhin2013,Lyutikov2021}.
This frequency characterises the linear acceleration emission.
It determines the modulational frequency for the free electron 
maser emission.
 
 \item An anomalous Doppler emission takes place when the resonance 
 condition $\omega(1 - n \frac{v_\parallel}{c} \cos \theta) + \omega_\mathrm{ce}/\gamma = 0$ is fulfilled, where 
 $s=-1$ is the resonance number for the cyclotron frequency $\omega_\mathrm{ce}$, $n$ is the refractive index, and $\theta$ is the propagation angle with respect to the magnetic field.
 This condition for the emission of X- or O-mode waves requires 
 $n > 0$ and $\gamma > v_A/c$, where $v_\mathrm{A}$ the Alfv\'{e}n speed, $c$ is the speed of light. 
 The emission frequency is $\omega = 2 (v_\mathrm{A}/c)^2 \omega_\mathrm{ce}/\gamma$, 
 which is in the radio band only for small values of $\omega_\mathrm{ce}/\gamma$ 
 \citep{Machabeli1979,Kazbegi1991,Lyutikov1999a,Lyutikov1999b}.
\end{itemize}

Note that so far scenarios C2 and C3 have been studied only by means of
oversimplifying approximations. The properties of the resulting
electromagnetic emissions are not well understood, yet.
In particular the consequences of density gradient effects 
are not taken into account, yet, in theoretical models trying to
theoretically elaborate scenarios C2 and 
C3~\citep{Weatherall1994,Rafat2019b,Rafat2019c,Manthei2021,Benacek2021}.
Also the bunch expansion, their interaction with the background plasma, and the 
nonlinear evolution after the overlap have not been considered, yet \citep{Usov2002,Ursov1988,Melikidze2000,Rahaman2020,Melrose2020b}.

In this paper, we model the evolution of the bunches as particle clouds in scenarios 
C2 and C3; i.e., with an enhanced central density and with a smooth 
decrease of the density from their maxima to the level of the 
surrounding plasma.
The  plasma between bunches is also taken into account, with a low but finite density.
We study the expansion of such a bunch, the propagation through the low density 
background, and for some cases, the eventual overlap of particles of adjacent bunches 
in phase space, the consequent growth of instabilities, the formation and evolution 
of waves.
In particular we investigate the evolution of the bunches and waves for three 
sets of characteristic initial parameters: 
the plasma temperature, the distance between bunches, 
and the relative drift speed of electrons and positrons.
We utilized one-dimensional (1D) PIC-code simulations, where the velocity distribution 
of only one finite component directed parallel to the magnetic field lines is
taken into account.
This approximation is valid for the lower pulsar magnetosphere, where the magnetic 
field is strong enough in order to immediately radiate away 
the energy of any perpendicular velocity component of the particles.

The paper is structured as follows: in Section~2 we describe the initial setup and 
numerical algorithms suitable for the kinetic study of relativistic pair plasmas
utilizing the appropriate PIC-code ACRONYM.
In Section~3 we present the resulting evolution of the bunches. 
First the main properties obtained in the simulations for all of them. 
We then, in more detail, discuss the results of the two most representative 
simulations.
An overall discussion of our results is contained in Section~4 and we state our
conclusions in Section~5.

%%%%%%%%%%%%%%%%%%%%%%%%%%%%%%%%%%%%%%%%%%%%%%%%%%%%%%%%%%
%%%%%%%%%%%%%%%%%%%%%%%%%%%%%%%%%%%%%%%%%%%%%%%%%%%%%%%%%%
\section{Methods} \label{sec:methods}
In order to numerically solve the kinetic Vlasov equation, we use the Particle-in-Cell (PIC) code ACRONYM.
The ACRONYM (“Another Code for Moving Relativistic Objects, 
Now with Yee lattice and
Macroparticles”) code \citep{Kilian2012} is a fully-kinetic
electromagnetic and explicit PIC code
that can handle ultrarelativistic particles.
We use its 1D3V version that is sufficient for the description of the relativistically hot plasma in the  lower pulsar magnetosphere, where the perpendicular motion of particles to the magnetic field lines can be neglected.
We use the Yee grid for the electric and magnetic fields \citep{Yee1966}, the standard Boris push \citep{Boris1970}, and the recently developed fourth-order shape factor
``Weighting with Time dependency'' WT4 \citep{Lu2020} to efficiently suppress
the numerical Cerenkov radiation.
The Cole-K\"ark\"ainen solver CK5 \citep{Karkkainen2006,Vay2011} is used 
to correctly propagate waves with phase velocity close to the speed of light. 
Moreover, current smoothing and periodic boundary conditions are applied.

\begin{table}
    \centering
    \begin{tabular}{llllrr}
    \hline \hline
    Sim. \#  & $\rho$ & $u_\mathrm{d}/c$ & $l/L$ & $l/d_\mathrm{e}$ & $\omega_\mathrm{p} t_\mathrm{end}$ \\
    \hline
    1 & 1    & 0   & 1/30 & 1200 & 15000 \\
    2 & 0.1  & 0   & 1/30 & 1200 & 15000 \\
    3 & 0.01 & 0   & 1/30 & 1200 & 15000 \\
    4 & 1    & 0   & 1/3  & 12000 & 15000 \\
    5 & 1    & 0.1 & 1/30 & 1200 & 5000 \\
    6 & 1    & 1   & 1/30 & 1200 & 5000 \\
    7 & 1    & 10  & 1/30 & 1200 & 5000 \\
    \hline \hline
    \end{tabular}
    \caption{List of simulations and their initial parameters: The simulation number, the initial inverse temperature $\rho$, the drift velocity $u_\mathrm{d}/c$, the relative distance between bunches $l/L$, the distance between bunches in units of skin depths $l/d_\mathrm{e}$, and the simulation time $\omega_\mathrm{p} t_\mathrm{end}$.}
    \label{tab:sim}
\end{table}

We run a series of simulations to sufficiently cover the
parameter space. Simulations are along the $x$-direction,
the direction of the magnetic field.
We use the time step $\omega_\mathrm{p} \Delta t = 0.025$
and the normalized grid cell size $\Delta x = 0.05\,d_\mathrm{e}$.
The plasma frequency of both species and 
the drift relativistic (Lorentz) factor are, respectively,
\begin{equation}
\omega_\mathrm{p} = \sqrt{\frac{n_0 e^2}{\gamma_\mathrm{d}m_\mathrm{e} \epsilon_0}}, \qquad \gamma_\mathrm{d} = \sqrt{1 + \frac{u_\mathrm{d}^2}{c^2}}.
\end{equation}
The particle density in the bunch center is $n_0$,
$u_\mathrm{d}$ is the initial drift velocity between electrons and positrons
associated with a particle drift momentum $p_\mathrm{d} = u_\mathrm{d}m_\mathrm{e}$,
$m_\mathrm{e}$ is the electron mass, $e$ is the electric charge, 
$\epsilon_0$ is the permittivity of vacuum.
The distance between bunches is $l = 1200\, d_\mathrm{e}$ and the ratio $L/l \approx 30$ were estimated by \citet{Usov2002}.
We set the plasma frequency to $\omega_\mathrm{p} = \sqrt{2}\times10^{10}$~s$^{-1}$ in all our simulations.
Each simulation has a physical length of $L = 720\,000 \Delta x$ ($36\,000\, d_\mathrm{e}$). The distance between bunches is $l = 1200\, d_\mathrm{e}$, except Simulation~4, where the distance is increased to $l = 12\,000\, d_\mathrm{e}$.
Simulations without an initial drift velocity run for $600\,000$ times steps
($\omega_\mathrm{p} t = 15\,000$), which is the time needed
for particles from the center of simulation box with a velocity close to the speed of light to approach the simulation boundary at $\sim (L-l) / (2 c) \approx 700\,000$ time steps.
Simulations with an initial drift velocity run for
200 000 time steps ($\omega_\mathrm{p} t = 5000$).
These simulations initially generate strong electric field that constrain particles.
The particle mixing eventually occurs, but without generation of a streaming instability in phase space.

Initially, we generate two particle species in the simulations --- electrons and positrons.
They obey a relativistically invariant Maxwell-J\"{u}ttner velocity distribution function
\begin{equation} \label{eq:distr}
 g(u) = \frac{1}{\gamma_\mathrm{d}}\frac{n}{2 K_1(\rho)} \mathrm{e}^{- \rho \gamma_\mathrm{d} \gamma (1 - \beta \beta_\mathrm{d})},
\end{equation}
where
\begin{equation}
 \gamma = \sqrt{1 + \frac{u^2}{c^2}}, \quad \beta = \frac{v}{c} = \frac{u}{c\gamma}, \quad \beta_\mathrm{d} = \sigma \frac{v_\mathrm{d}}{c} = \sigma\frac{u_\mathrm{d}}{c\gamma_\mathrm{d}},
\end{equation}
where $\rho = m_\mathrm{e}c^2 / k_\mathrm{B}T$ is the inverse temperature,
$k_\mathrm{B}$ is the Boltzmann constant,
$\sigma = 1$ for electrons, $\sigma = -1$ for positrons,
$K_1$ is the MacDonald function of the first order (modified Bessel function of
the second kind).
Particles are generated by using the rejection method that
selects particles from a velocity interval that is much wider than
the expected distribution width.
The produced distributions converge to the analytical prescription 
as the number of particles increases.
$x = 0$ is located exactly between bunches at the center of our simulation box.
The initial particle density $n(x)$ is
\begin{equation}
  n(x) =   
 \left\{
 \begin{array}{ll}
  0.1 \, n_0, & |x| \leq \frac{l}{2}, \\
  n_0 \, \mathrm{exp}\left\{ - \frac{(\frac{L}{2} - |x|)^6}{x_0} \right\}     & |x| > \frac{l}{2}, \\
 \end{array}
 \right.
\end{equation}
\begin{equation}
 x_0 = \left( - \ln(0.1) \right)^{-\frac{1}{6}} \frac{L - l}{2}  \approx 0.870\cdot \frac{L - l}{2}
\end{equation}
where $x_0$ is chosen $n(x)$ be a smooth function in $x = \pm l/2$.
We assume that the initial plasma density is an even function.
The simulation part where $x< -l/2$ covers half of the trailing bunch,
while $x > l/2$ covers half of the leading bunch.
We initially have the same number of electrons and positrons in each grid cell.
The net-current is subtracted in each time step.
In all simulations we represent the density $n_0$ by $10^4$ macro-particles per cell.
The Debye length for a relativistically hot plasma with a 1D velocity function is given by \citep{Diver2015}
\begin{equation}
 \lambda_\mathrm{D} = \frac{c_s^r}{\omega_\mathrm{p}} = \frac{c}{\omega_\mathrm{p}} \left[ \frac{1}{G_\mathrm{1D}}\frac{\mathrm{d}G_\mathrm{1D}}{\mathrm{d}\rho} \left( \rho \frac{\mathrm{d}G_\mathrm{1D}}{\mathrm{d}\rho} + \frac{1}{\rho}  \right)^{-1} \right]^{\frac{1}{2}}
\end{equation}
\begin{equation}
G_\mathrm{1D}(\rho_{0,1}) = \frac{K_2(\rho_{0,1})}{K_1(\rho_{0,1})},
\end{equation}
where $c_s^r$ is the relativistic ``sound'' speed.
In such a plasma, particles have 2 degrees of freedom
(one for electron and one for positron)
and the adiabatic index is $4/2$.
For $\rho \rightarrow 0$, the relativistic sound speed approaches the speed of light.
For $\rho = 1 - 0.01$, the ratio of the Debye length with respect to the grid size ranges from $\lambda_\mathrm{D}/\Delta x = 18.2-20.0$
at the bunch center to $\lambda_\mathrm{D}/\Delta x = 57-63$ between the bunches.
Thus, the number of particles in a Debye length is $(1.8-6.3)\times 10^5$.

The initial electric field is zero.
The magnetic field is not included in the simulation,
as the particle distribution is only  1-dimensional and all quantities vary only in the $x$-direction and do not interact with the magnetic field.
The currents and evolution of the electric field can be only along this direction.
Then, the $curl$ operator acting on electric and magnetic fields is always zero in the Maxwell equations and
the electric and magnetic fields do not influence each other.
The parameters of the different simulation runs are summarized in Table~\ref{tab:sim}.
The initial values range as follows: inverse temperature range $\rho = [1,0.01]$, 
distance between bunches range $l/L = [1/30,1/3]$, and drift velocity range $u_\mathrm{d}/c = [0,10]$.

%%%%%%%%%%%%%%%%%%%%%%%%%%%%%%%%%%%%%%%%%%%%%%%%%%%%%%%%%%
%%%%%%%%%%%%%%%%%%%%%%%%%%%%%%%%%%%%%%%%%%%%%%%%%%%%%%%%%%
\section{Results} \label{sec:results}
We denote the bunch in $x < 0$ as the ``left bunch'',
the bunch in $x > 0$ as the ``right bunch'', and the region around $x \approx 0$ as the ``center''.
All processes are described and presented in the simulation reference frame that is moving with respect to the pulsar reference frame.

\begin{figure*}[!ht]
    \centering
    \includegraphics[width=0.32\textwidth]{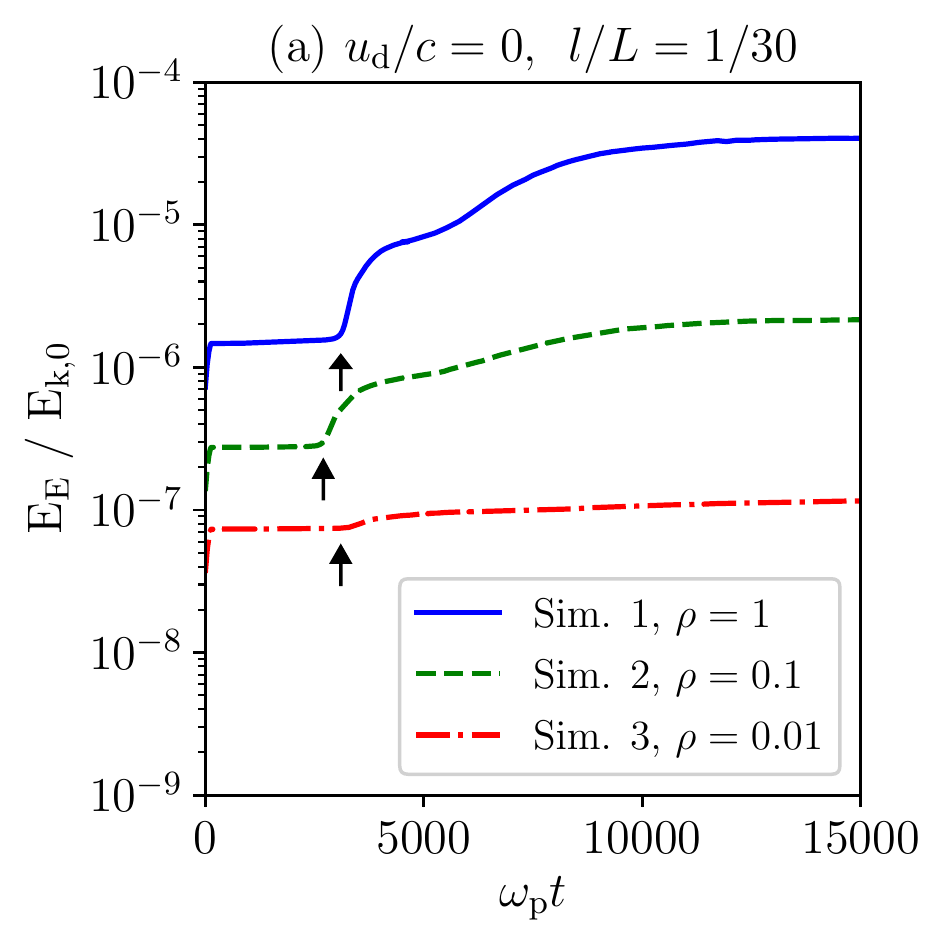}
    \includegraphics[width=0.32\textwidth]{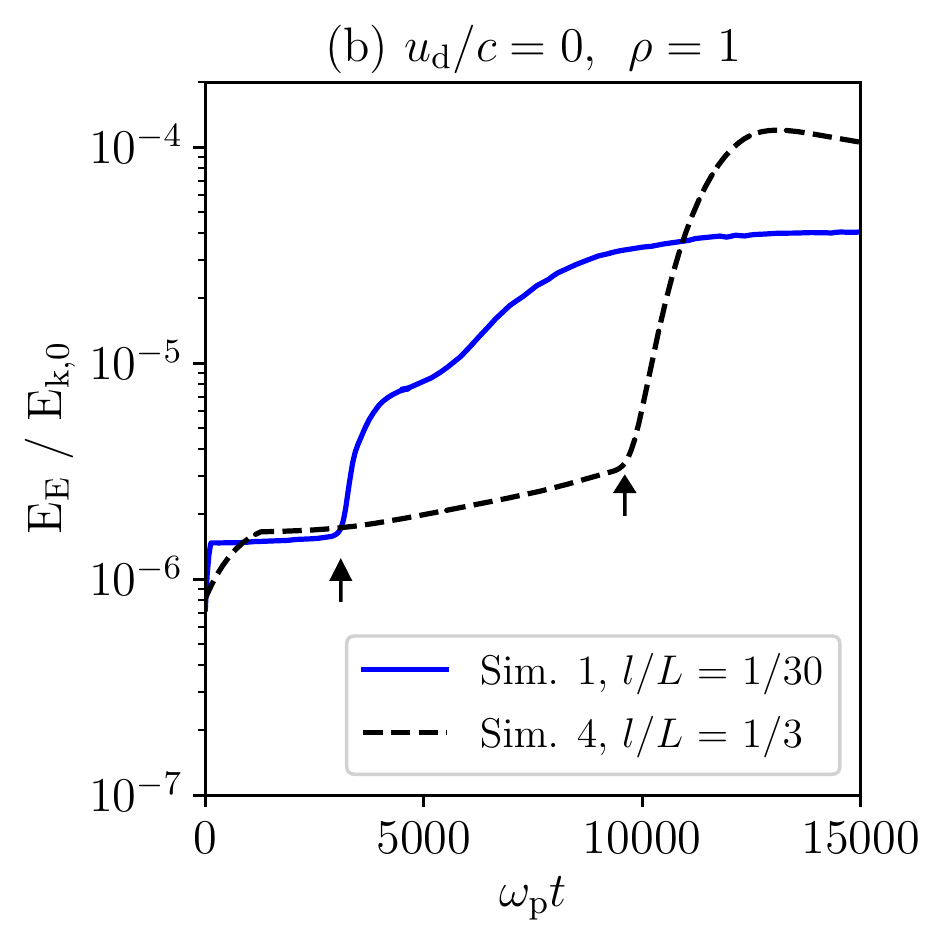}
    \includegraphics[width=0.32\textwidth]{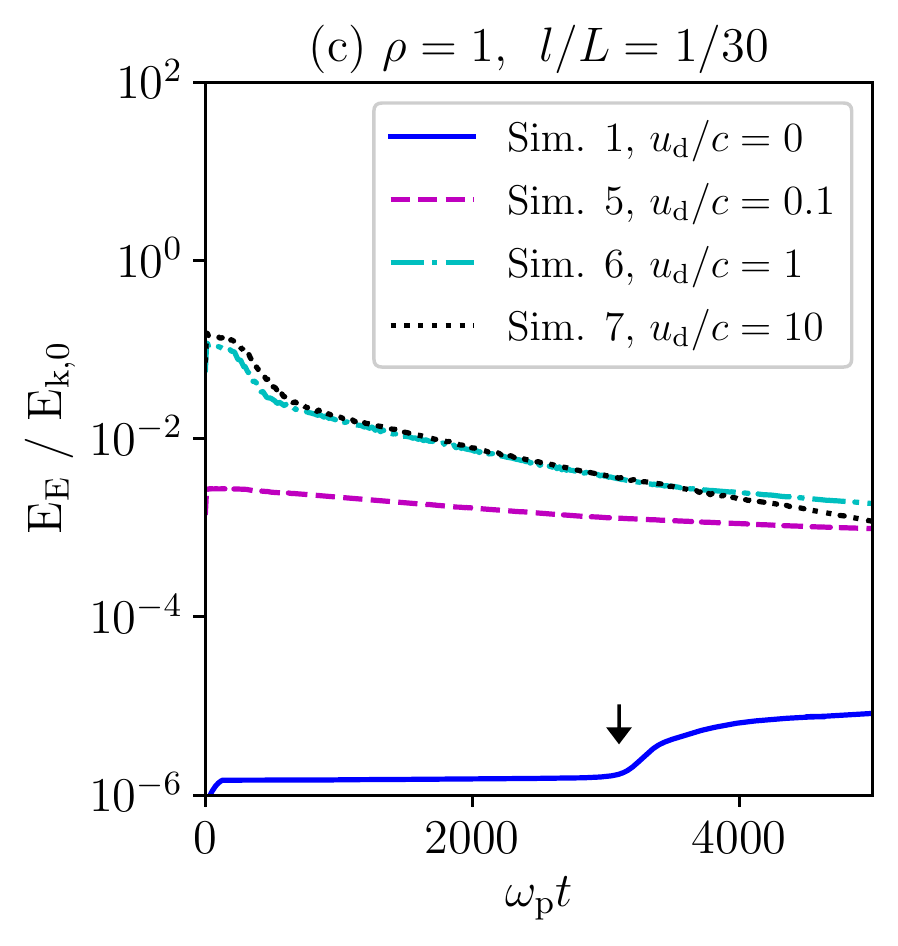}
    \caption{Evolution of the electric to kinetic energy ratio for simulations with different: initial inverse temperatures (a), initial distance between bunches (b), initial drift velocity between species (c). Scales of axes differs between subfigures. 
    Black arrows indicate the time when the bunch interaction begins in Simulations~1-4.}
    \label{fig:energyevolution}
\end{figure*}

\subsection{Evolution of Simulations}
\subsubsection{Energy Evolution}
Figure~\ref{fig:energyevolution} shows the evolution of the ratio of the electric
to the initial total kinetic energy for all simulations.
The evolution of Simulation~1
is plotted as a blue line in all subfigures as the reference. 
The plasma inverse temperature, distance between bunches,
and drift between species are varied between the different simulations.

In Simulations~1-4 with zero initial drift between species (Figure~\ref{fig:energyevolution}a,b),
particles from both bunches begin to expand after the simulation start.
The expansion is not suppressed by ambipolar diffusion
as in the case of an electron-ion plasma.
Both particle species have the same initial velocity distribution 
and both expand in the same way due to the same inertia of electrons and positrons.
The fastest particles from the left bunch reach the
fastest particles from the right bunch in the simulation center,
and they cause a two-stream instability.
The start of the mutual particle interaction is denoted by a black arrow for Simulations~1-4.
By using the electric field in the whole simulation domain,
its growth rates $\Gamma / \omega_\mathrm{p}$ are
$(3.48 \pm 0.05) \times 10^{-3}$ for Simulation~1 in the time interval $\omega_\mathrm{p}t = [2082,3390]$,
$(3.35 \pm 0.04) \times 10^{-3}$ for Simulation~2 in the time interval $\omega_\mathrm{p}t = [2060,2880]$, and
$(5.8 \pm 0.4) \times 10^{-4}$ for Simulation~3 in the time interval $\omega_\mathrm{p}t = [2610,3650]$.
With increasing plasma temperature,
the instability growth rates are decreasing,
and the amount of converted kinetic energy into electrostatic energy also decreases.
In the hottest case,
the energy ratio increases only slightly above the initial energy level.
The evolution slightly differs in Simulation~4,
where the instability produced by the interaction of expanding particles with the background plasma has enough time to develop.
The growth rates of this instability is $\Gamma / \omega_\mathrm{p} = (7.8 \pm 0.5) \times 10^{-5}$ in the time interval $\omega_\mathrm{p}t = [2920,10150]$.
Moreover, particles from both bunches need more time to approach each other
during the bunch expansion.
That occurs not only due to the larger bunch separation but also due to the slow down of particles by the developed streaming instability.
The streaming instability in the center has a lower growth rate  $(2.69 \pm 0.02) \times 10^{-3}$ in the time interval $\omega_\mathrm{p}t = [10100,10980]$.

In Simulations~5-7 with non-zero initial drift velocity (Figure~\ref{fig:energyevolution}c),
the simulation evolution is different.
At the density gradient of the bunches, particles quickly form a local nonzero charge density and ambipolar diffusion that generates strong electric fields.
Even a small drift velocity $u_\mathrm{d}/c = 0.1 < 1/\sqrt{\rho}$ increases
the amount of the released energy by $\sim 3$ orders of magnitude in Simulation~5.
As the drift velocity increases in Simulations~6 and 7, the amount of converted energy approaches $\sim 0.15\,\%$ of the total initial kinetic energy.
In all three Simulations~5-7, the electrostatic energy gradually decreases after short time and then the system evolves with a constant energy ratio.

\subsubsection{Spatial Evolution along Simulation Domain}
\begin{figure*}[!ht]
    \centering
    \includegraphics[width=\textwidth]{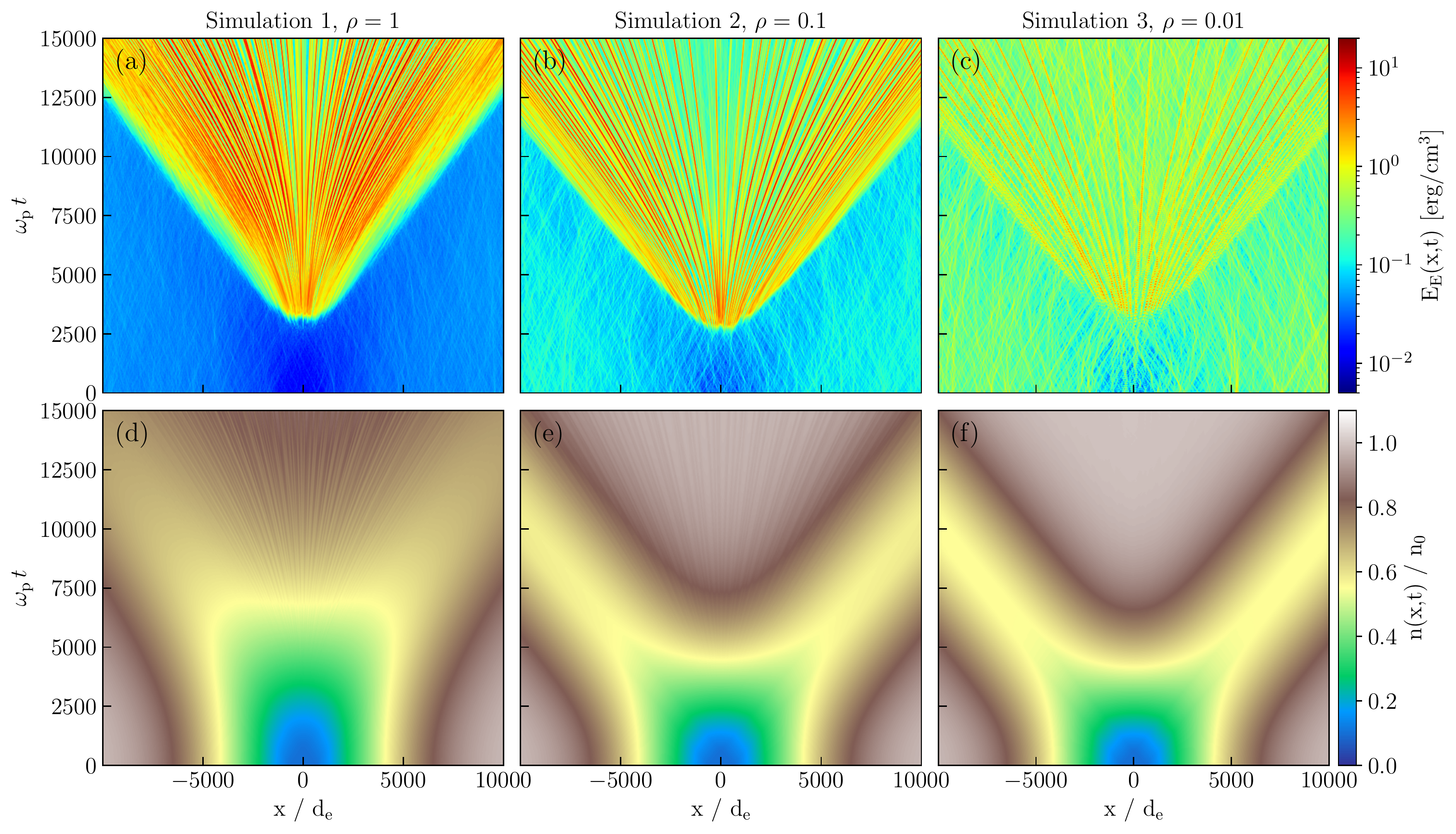}
    \vspace{-0.3cm}
    \caption{Evolution of the electrostatic energy density (a-c) and particle density (d-f) for inverse temperatures $\rho = [1,0.01]$ along $x$-direction.
    Simulation~1 (a,d), Simulation~2 (b,e), Simulation~3 (c,f). In all cases $u_\mathrm{d}/c=0$, $l/L = 1/30$.
    }
    \label{fig:spatialevolution1}
\end{figure*}

\begin{figure}[!ht]
    \centering
    \includegraphics[width=0.45\textwidth]{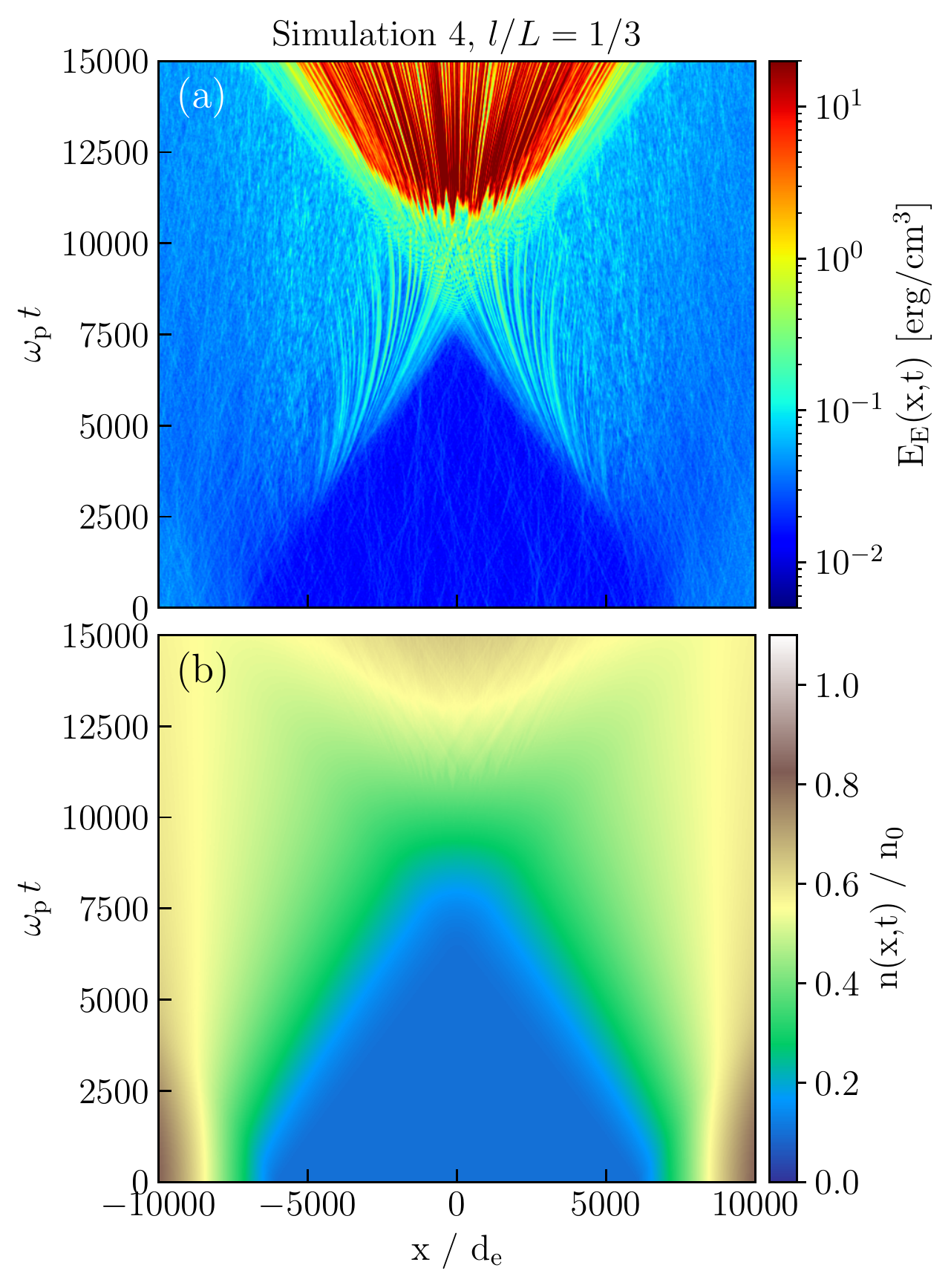}
    \vspace{-0.3cm}
    \caption{Evolution of the electrostatic energy density (a) and particle density (b) for distance between bunches $l/L = 1/3$, $\rho = 1$, $u_\mathrm{d}/c=0$ along $x$-direction in Simulation~4. Compare with Figure~\ref{fig:spatialevolution1}.
    }
    \label{fig:spatialevolution2}
\end{figure}

\begin{figure*}[!ht]
    \centering
    \includegraphics[width=\textwidth]{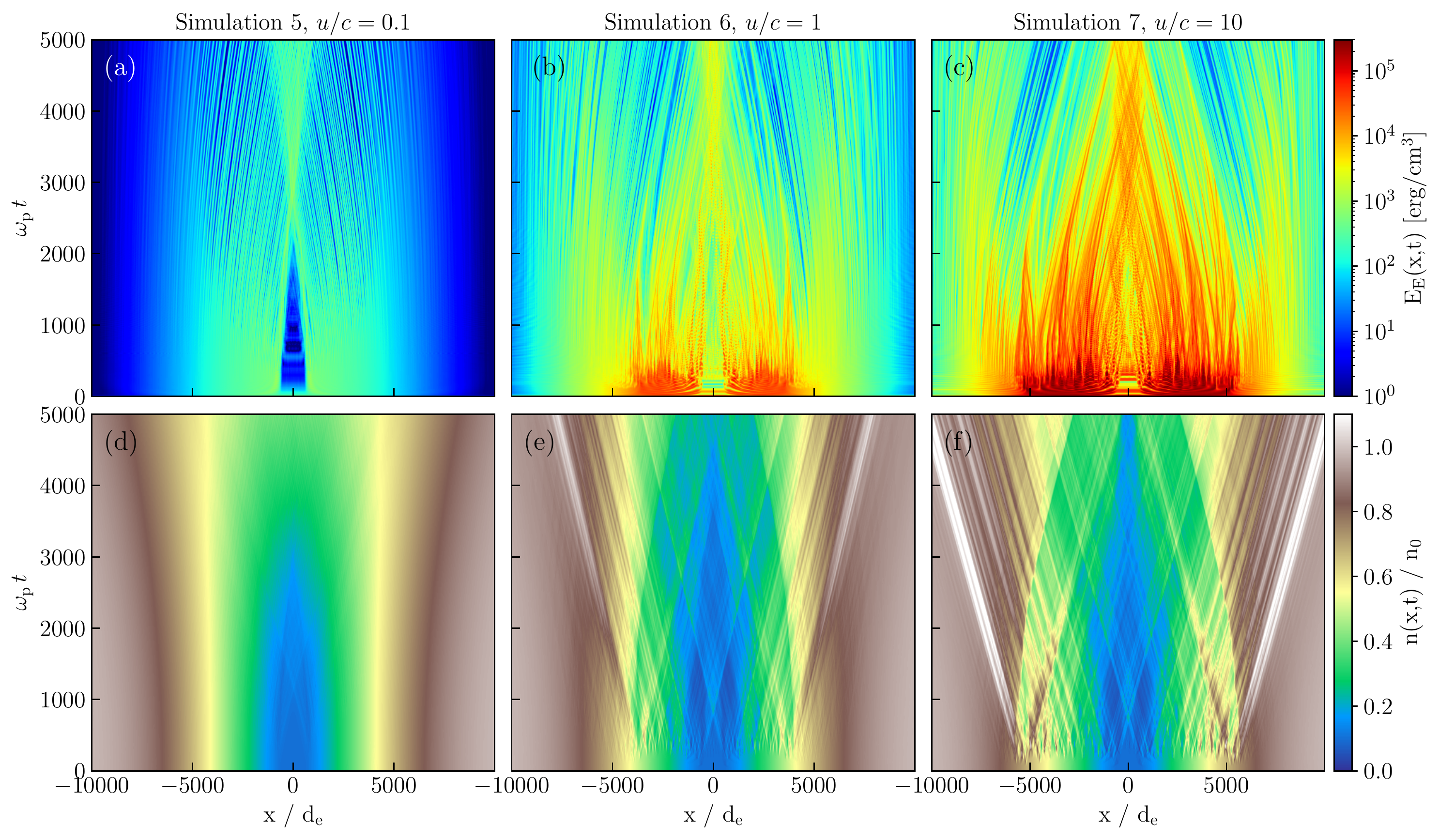}
    \vspace{-0.3cm}
    \caption{Evolution of electrostatic energy density (a-c) and electron density (d-f) for initial drift velocities $u_\mathrm{d}/c=[0.1,10]$ along $x$-direction.
    Simulation~5 (a,d), Simulation~6 (b,e), Simulation~7 (c,f).
    In all cases $\rho = 1$, $l/L = 1/30$. Compare with Figures~\ref{fig:spatialevolution1} and \ref{fig:spatialevolution2}.
    }
    \label{fig:spatialevolution3}
\end{figure*}

In Figures~\ref{fig:spatialevolution1}--\ref{fig:spatialevolution3},
we present the evolution of the electrostatic energy density
and electron density, which is normalized to the initial density at the bunch center $n_0$, along the simulation box.

Figure~\ref{fig:spatialevolution1} shows the evolution for Simulations~1-3.
As particles from the bunches expand, the electrostatic energy forms a visible ``web'' structure of waves in the background electric field (blue regions).
When particles begin to interact in the simulation center at $\omega_\mathrm{p} t \sim 3000$,
electrostatic waves are formed at time $\sim 4000\, \omega_\mathrm{p} t$,
which are stable for long time.
Electrostatic waves generated in $x<0$ start propagating to the left, while those in $x>0$ to the right, respectively.
The distance between generated waves increases in time.
New waves are formed at the edges of expanding bunches.
With increasing temperature, the distance between them increases
while the intensity of their electric fields decreases.
The total particle density (Figure~\ref{fig:spatialevolution1}d-f)
decreases in the centers of the bunches as they expand,
and the low density increases in the center.

Figure~\ref{fig:spatialevolution2} shows the effect of increasing distance between bunches in Simulation~4.
Waves generated by the streaming instability have higher energies
and smaller separations between them.
Furthermore, the density fluctuations are smaller than in Simulation~1.

Figure~\ref{fig:spatialevolution3} illustrates the effects of the initial drift velocity in Simulations~5-7.
Please note that these figures have different color schemes
than Figures~\ref{fig:spatialevolution1}-\ref{fig:spatialevolution2}.
The mean intensity of electrostatic waves is much higher in these three cases.
The most intense electrostatic waves are created at regions
with the largest density gradients.
Until $\omega_\mathrm{p}t \sim 1000$,
the group velocity of these waves is close to zero.
Then, it increases towards the simulation center,
where they eventually interact and may be absorbed.
In these three simulations, the density evolution does not fully correspond
to the evolution of the electric field.
While for $u_\mathrm{d}/c = 0.1$, the density gradient evolves more or less smoothly,
with increasing drift velocity strong density waves are formed.
Moreover, the propagation direction of the particle density waves is not
only toward the simulation center, but also outwards.

\begin{table*}
    \centering
    \begin{tabular}{crc|lll|lll}
    \hline \hline
    ~ & ~ & ~ & \multicolumn{3}{c}{Maximum} &  \multicolumn{3}{c}{Mean} \\
    Sim. \# & $\omega_\mathrm{p} t$ & $x$ & $E_\mathrm{E} / E_\mathrm{k,0}$ & $E_\mathrm{E}$ & $|E|$ & $E_\mathrm{E} / E_\mathrm{k,0}$ & $E_\mathrm{E}$ & $|E|$ \\
    ~ & ~ & [$d_\mathrm{e}$] & ~ & [erg/cm$^3$] & [V/cm] & ~ & [erg/cm$^3$] & [V/cm]\\
    \hline
    1 & 15000 & $0-12500$ & $1.7\times10^{-3}$ & $1.3\times10^{2}$ & $ 1.7\times 10^{4}$ & $2.2\times10^{-5}$ &   $1.5\times10^{0}$ & $ 1.2\times 10^{3}$\\
    2 & 15000 & $0-14000$ & $1.4\times10^{-4}$ & $6.1\times10^{1}$ & $ 1.2\times 10^{4}$ & $2.3\times10^{-6}$ & $1.0\times10^{0}$ & $ 9.8\times 10^{2}$ \\
    3 & 15000 & $0-14500$ & $3.3\times10^{-6}$ & $1.4\times10^{1}$ & $ 5.6\times 10^{3}$ & $1.3\times10^{-7}$ & $5.5\times10^{-1}$ & $ 8.6\times 10^{2}$ \\
    4 & 15000 & $0-6500$  & $1.1\times10^{-3}$ & $1.6\times10^{2}$ & $1.9 \times 10^{4}$ & $9.3\times10^{-5}$ & $5.8\times10^{0}$ & $2.3 \times 10^{3}$ \\
    5 & 1000  & $0-9500$ & $8.0\times10^{-1}$ & $1.7\times10^{3}$ & $ 6.2\times 10^{4}$ & $4.3\times10^{-3}$ & $1.5\times10^{2}$ & $ 1.3\times 10^{4}$ \\
    6 & 1000  & $0-9500$ & $1.1\times10^{0}$ & $4.9\times10^{4}$ & $ 3.3\times 10^{5}$ & $1.8\times10^{-2}$ & $1.9\times10^{3}$ & $ 4.5\times 10^{4}$ \\
    7 & 1000  & $0-9500$ & $1.1\times10^{0}$ & $2.5\times10^{5}$ & $ 7.5\times 10^{5}$ & $2.3\times10^{-2}$ & $1.0\times10^{4}$ & $ 1.1\times 10^{5}$ \\
    \hline \hline
    \end{tabular}
    \caption{Maximal and mean values of  the electric field component found in the range along $x$ and time $\omega_\mathrm{p} t$:
    The ratio of the electric and kinetic energy density $E_\mathrm{E}/E_\mathrm{k,0}$,
    the electrostatic energy density $E_\mathrm{E}$,
    and absolute value of the electric intensity $|E|$.}
    \label{tab:density}
\end{table*}

Table~\ref{tab:density} summarizes the properties of the electric field intensity in all simulations.
We select the final time $15000\,\omega_\mathrm{p} t$
to analyze the simulations with $u_\mathrm{d}/c = 0$,
and time $1000\,\omega_\mathrm{p} t$ for simulations
with $u_\mathrm{d}/c \neq 0$.
The spatial region is selected to cover the region of generated waves in all cases. 
Maximum, as well as mean values, are calculated in the selected regions.
Note that $E_\mathrm{E}(x,t)$ is electrostatic energy density varying with position and time,
while $E_\mathrm{k,0}(t=0)$ (Figure~\ref{fig:energyevolution}) is the mean kinetic energy density at the simulation start.

\subsubsection{Evolution in Fourier Space}
\begin{figure*}[!ht]
    \centering
    \includegraphics[width=\textwidth]{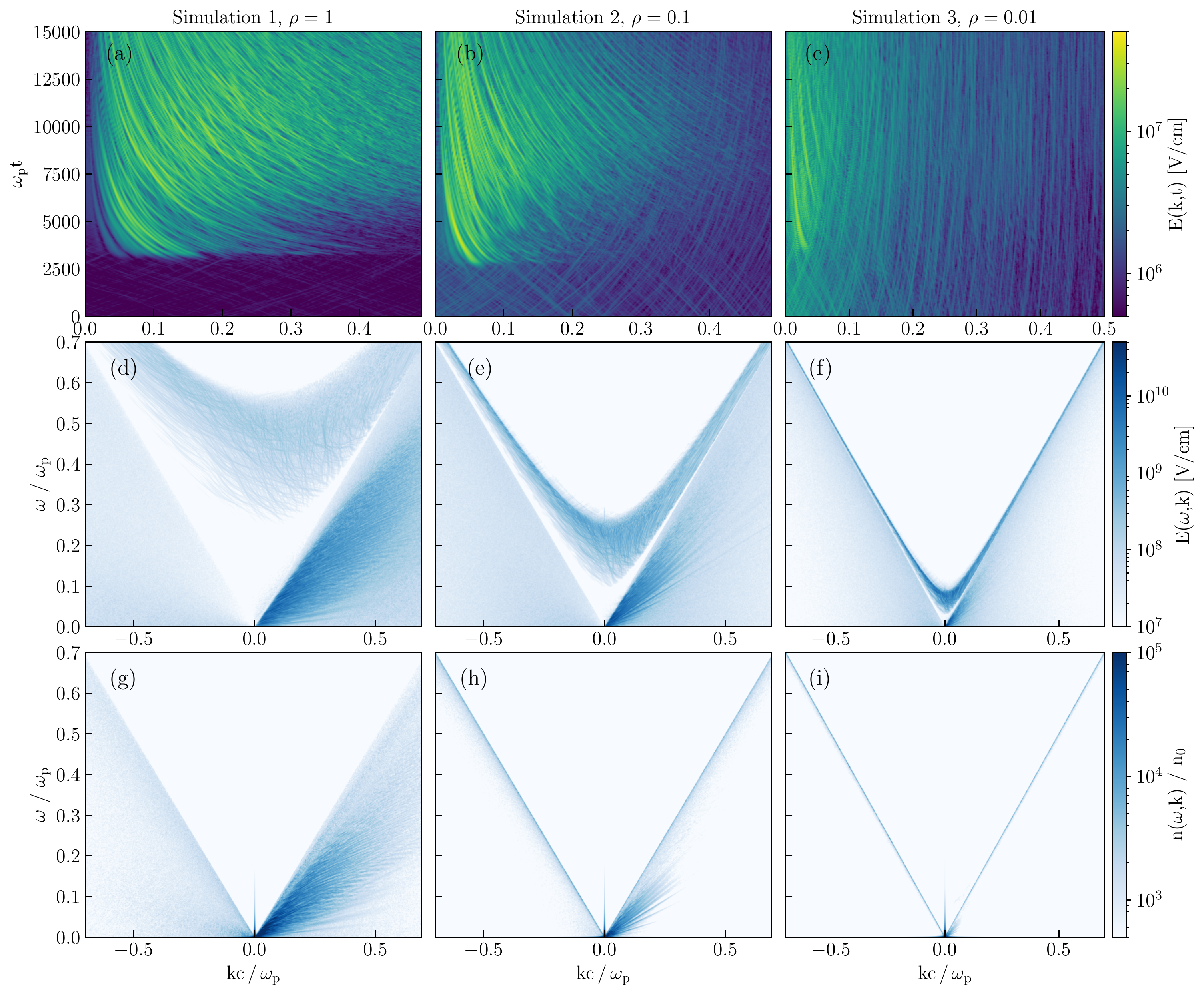}
    \vspace{-0.3cm}
    \caption{Evolution of the electrostatic waves as a function of the wavenumber (a-c),
    dispersion diagrams of the electrostatic waves (d-f),
    and dispersion diagrams of the particle density waves (g-i)
    for the inverse temperatures $\rho = [1,0.01]$ along $x$-direction.
    All quantities are selected in $x / d_\mathrm{e} = [0,6000]$ and the whole simulation time.
    Simulation~1 (a,d,g), Simulation~2 (b,e,h), Simulation~3 (c,f,i),
    all cases $u_\mathrm{d}/c=0$, $l/L = 1/30$.}
    \label{fig:fourierevolution1}
\end{figure*}

\begin{figure}[!ht]
    \centering
    \includegraphics[width=0.45\textwidth]{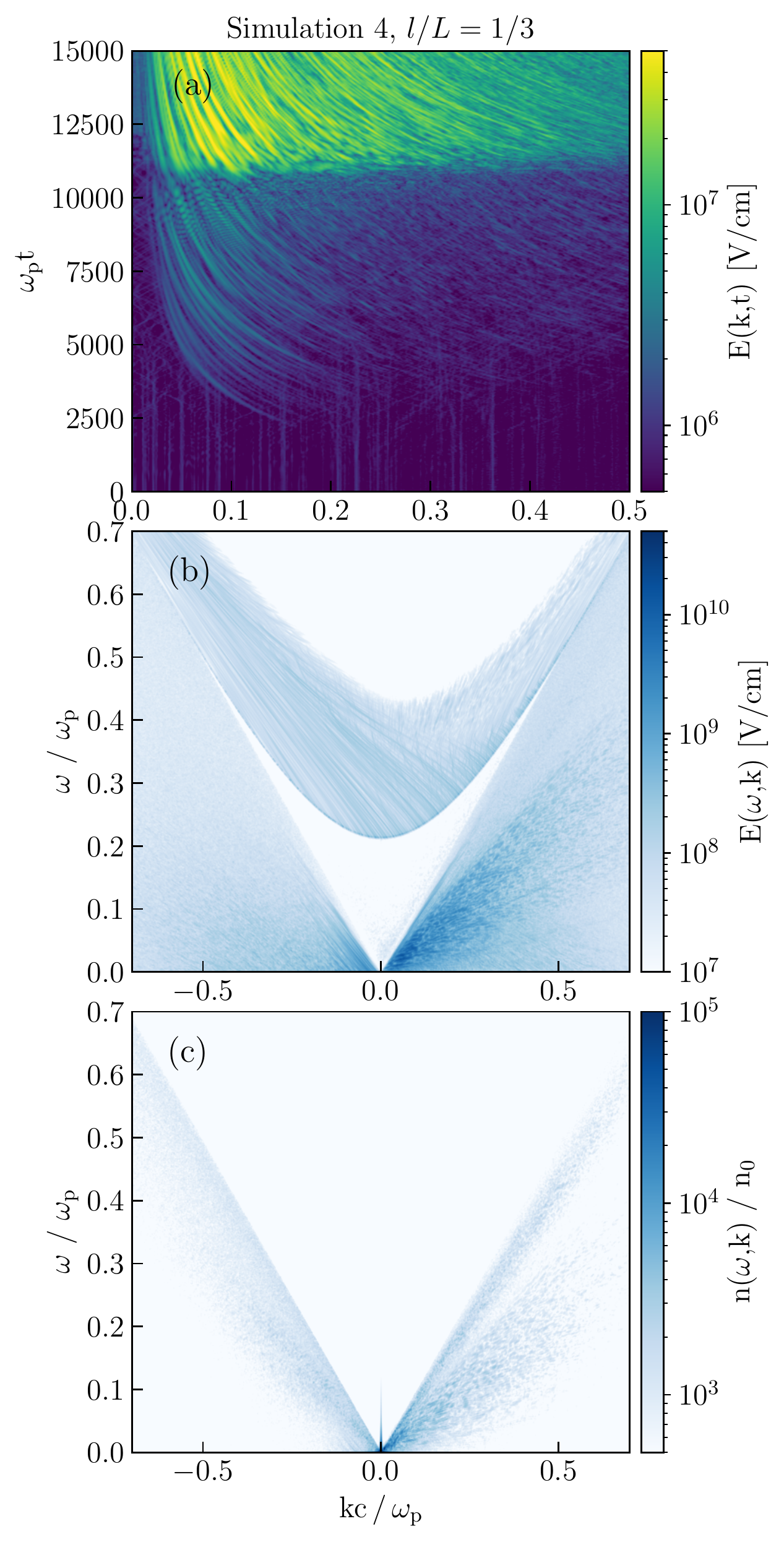}
    \vspace{-0.3cm}
    \caption{Evolution of the electrostatic waves as a function of the wavenumber (a),
    dispersion diagram of the electrostatic waves (b),
    and dispersion diagram of the particle density waves (c)
    for the initial distance between bunches $l/L = 1/3$, $\rho = 1$, $u_\mathrm{d}/c=0$ along $x$-direction in Simulation~4.
    All quantities are selected in $x / d_\mathrm{e} = [0,6000]$ and the whole simulation time.
    Compare with Figure~\ref{fig:fourierevolution1}.}
    \label{fig:fourierevolution2}
\end{figure}

\begin{figure*}[!ht]
    \centering
    \includegraphics[width=\textwidth]{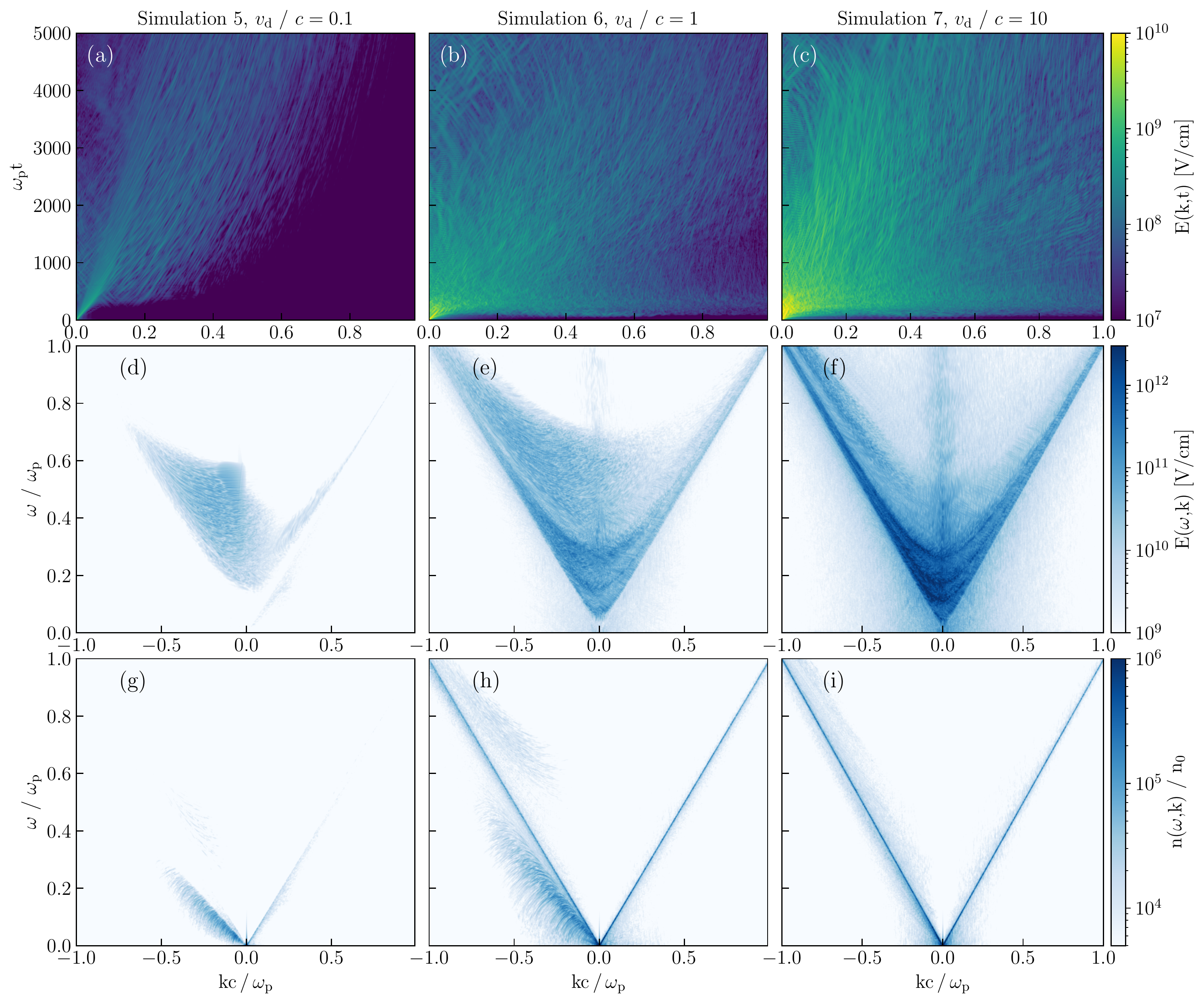}
    \vspace{-0.3cm}
    \caption{Evolution of the electrostatic waves as a function of the wavenumber (a-c),
    dispersion diagrams of the electrostatic waves (d-f),
    and the dispersion diagrams of the particle density waves (g-i)
    for the initial drift velocity $u_\mathrm{d}/c= [0.1,10]$ along $x$-direction.
    All quantities selected in $x / d_\mathrm{e} = [0,6000]$ and the whole simulation time.
    Simulation~5 (a,d,g), Simulation~6 (b,e,h), Simulation~7 (c,f,i).
    In all cases $\rho = 1$, $l/L = 1/30$.
    Compare with Figures~\ref{fig:fourierevolution1} and~\ref{fig:fourierevolution2}.
    }
    \label{fig:fourierevolution3}
\end{figure*}

Figures~\ref{fig:fourierevolution1}-\ref{fig:fourierevolution3} show the evolution
of the electrostatic wavenumber, dispersion diagrams of the electrostatic waves,
and the dispersion diagrams of the density waves.
In order to build the dispersion diagrams, the region $x/d_\mathrm{e} = [0,6000]$ and the whole simulation time every 120th timestep are considered.

The evolution of the wavenumber as a function of temperature is presented in Figure~\ref{fig:fourierevolution1}a-c for Simulations~1-3. 
During the bunch expansion $\omega_\mathrm{p}t < 3000$,
low amplitude electrostatic waves are present as
pairs of counter-propagating waves, which are initially created at the same wavenumber.
Their propagation speed decreases with increasing plasma temperature.
When particles from bunches begin to interact $\omega_\mathrm{p}t \sim 3000$,
only electrostatic waves are created (green-yellow lines).
The number of generated waves and their wavelengths decrease with
increasing plasma temperature.
In addition, the wavenumber of these waves decreases in time and additional waves are formed at larger wavelengths.

The dispersion diagrams of the electrostatic waves (Figure~\ref{fig:fourierevolution1}d-f)
shows the generation of two types of waves.
The superluminal L-mode wave and the diagonal mode with subluminal L-mode waves.
The frequencies of both superluminal and subluminal waves
decrease with increasing the temperature
due to dispersion effects in the relativistically hot plasma rest frame
(see e.g., \cite{Rafat2019a}, Figure~7) as $\langle \gamma^{-3} \rangle$.
The superluminal L-modes waves are strongly broadened in frequency due to the fact
that each dispersion diagram takes data over a broad range of plasma densities.
Moreover, they exhibit an internal fine structure and their intensity increases with increasing temperature.
There is only a limited number of the subluminal L-mode waves that 
are composed of modes with different group velocities.

The density waves (Figure~\ref{fig:fourierevolution1}g-i) do not form superluminal L-mode waves because there is no part of the distribution function that could support superluminal density waves.
The density wave waves are constrained to subluminal electrostatic waves.
Moreover, the density waves with $v_\phi \approx c$ are stronger for higher plasma temperatures.

When bunches are separated by a larger distance in Simulation~4 (Figure~\ref{fig:fourierevolution2}),
the wavenumbers of electrostatic waves manifest a slightly different behavior.
The initial ($\omega_\mathrm{p}t < 3000$) low amplitude background waves are not created in pairs.
Instead, only waves with constant wavenumber in time are present. 
As the expanding particles from the bunches interact with the background $3000 < \omega_\mathrm{p}t < 10500$, they form beam waves.
Finally ($\omega_\mathrm{p}t > 10500$) the most intense waves are created during the interaction of particles from both bunches.
The wavenumber of both medium and high intensity waves decreases in time.
The dispersion of electrostatic waves and electron density waves is similar to
Simulation~1 (Figure~\ref{fig:spatialevolution1}d,g).
Small differences are that the superluminal L-mode waves form clearly visible ``striations'' 
in the ranges $\omega/\omega_\mathrm{p} \sim [0.3,0.6]$ and $kc/\omega_\mathrm{p} \sim [-0.5,0]$.
Subluminal waves with $\omega/\omega_\mathrm{p} = [0,0.1]$
and $kc/\omega_\mathrm{p} = [-0.1,0]$ are enhanced,
and the subluminal waves with positive $k$ do not form contiguous modes.

Figures~\ref{fig:fourierevolution3}a-c (for Simulations~5-7) show the evolution of the wavenumber for increasing initial drift velocity.
The waves are created at $kc/\omega_\mathrm{p} \sim 0$ right after the simulation start;
their wavenumbers increase in time.
The dispersion of electrostatic waves (Figures~\ref{fig:fourierevolution3}d-f)
mostly manifest superluminal L-mode waves with negative wavenumbers.
The intensity of waves increases with increasing drift velocity.
Low intensity subluminal L-mode waves are enhanced with $v_\phi < c$ and positive wavenumber.
The density dispersion (Figures~\ref{fig:fourierevolution3}g-i)
shows subluminal waves with negative wavenumber and increasing intensity of waves $v_\phi \approx c$ with increasing temperature.

%%%%%%%%%%%%%%%%%%%%%%%%%%%%%%%%%%%%%%%%%%%%%%%%%%%%%%%%%%%%%%%%%%%%%%%%%%%%%%%%
\subsection{Effects of the Initial Drift Velocity}
In this section, we present detailed results of Simulations~1 and 6.
In Simulation~1, the initial drift is zero, in Simulation~6,
the initial drift between electrons and positrons equals to the thermal velocity.

\subsubsection{Evolution of the Electron Phase Space}
\begin{figure*}[!ht]
    \centering
    \includegraphics[width=0.49\textwidth]{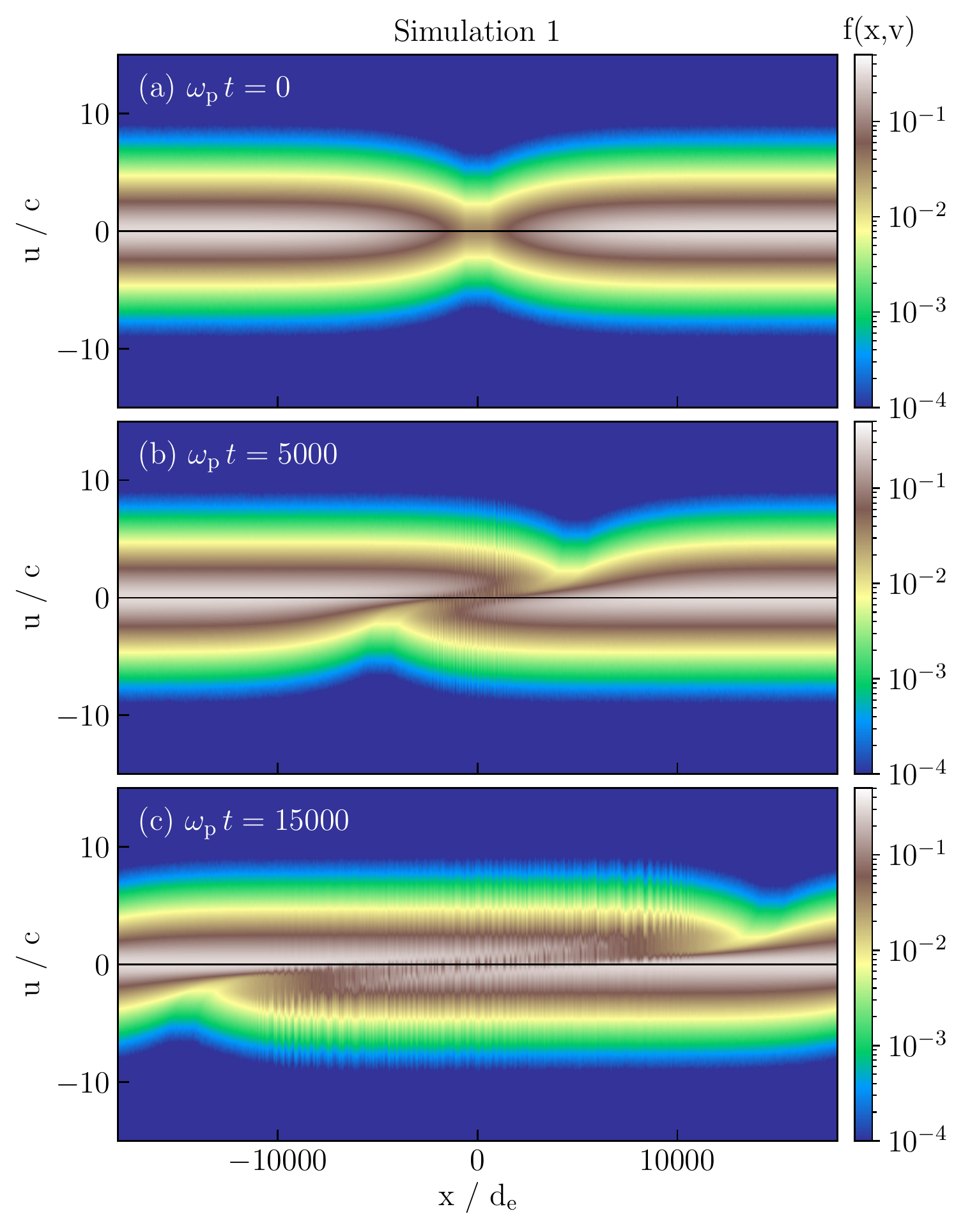}
    \includegraphics[width=0.49\textwidth]{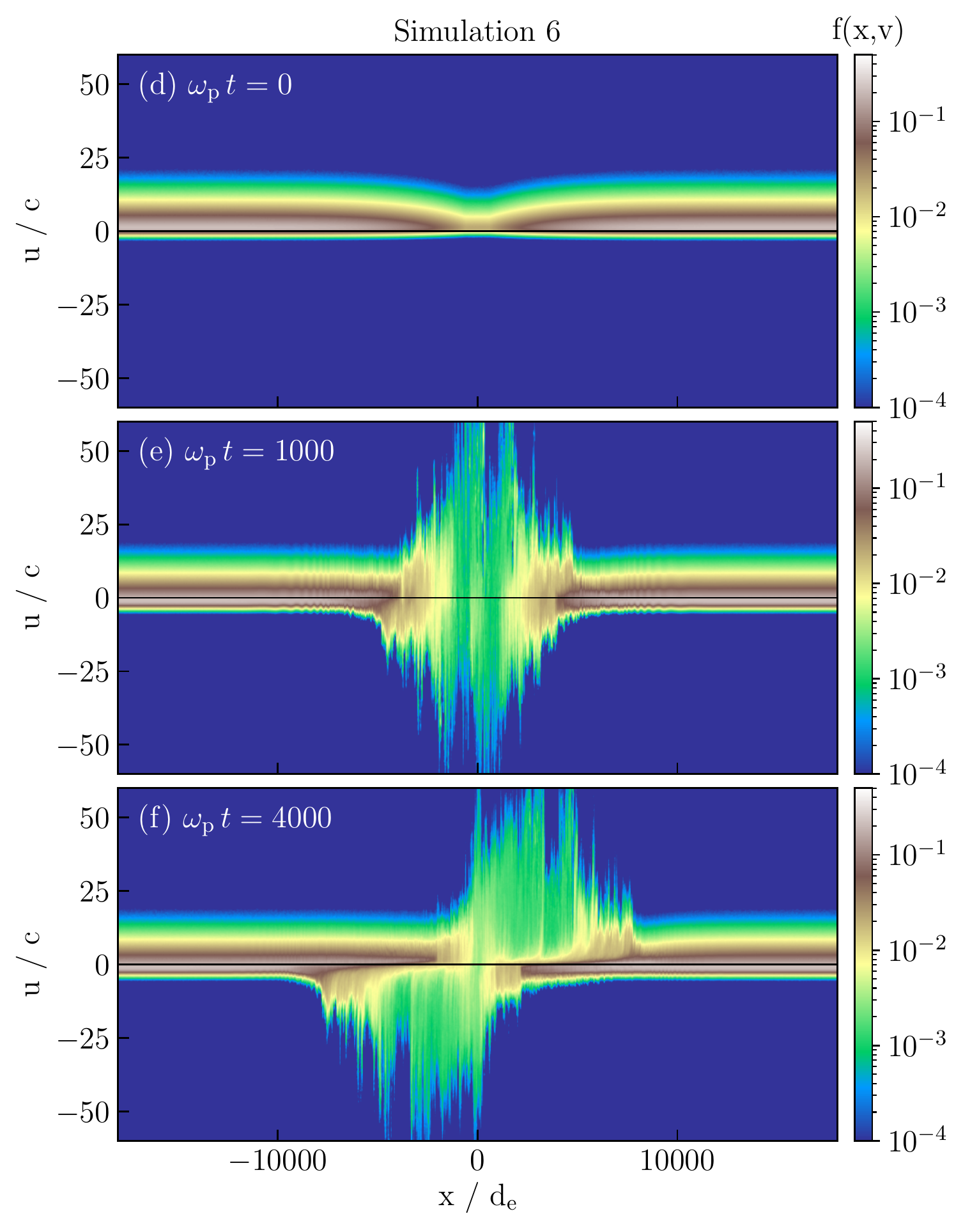}
    \vspace{-0.3cm}
    \caption{
    Electron phase space evolution of Simulation~1 (a-c) and Simulation~6 (d-e).
    Each case for three selected time moments. Note that the shown range of velocities is larger in Simulation~6.
    }
    \label{fig:phasespace}
\end{figure*}

Figure~\ref{fig:phasespace} shows the electron phase spaces of the whole domain,
for three selected time steps.
As particles overlap in phase space in Simulation~1,
they begin to form instability and waves.
The electrostatic waves influence the slower particles with $|u|/c < 2$
and form phase space holes.
In Simulation~6, the shown range of velocities is larger.
Plasma heats up in the regions with the strongest electrostatic waves (Figure~\ref{fig:phasespace}e,f).
The tail of the distribution can exceed $|u|/c \sim 50$.
For $x > 0$, particles have mostly positive velocities,
while they have mostly negative ones for $x < 0$.
Their distributions are not symmetric with respect to the axis $u = 0$, and generally, they are stable with respect to the streaming instability formed in Simulation~1.

~

\subsubsection{Evolution of Electrostatic Energy Density}
\begin{figure*}[!ht]
    \centering
    \includegraphics[width=0.49\textwidth]{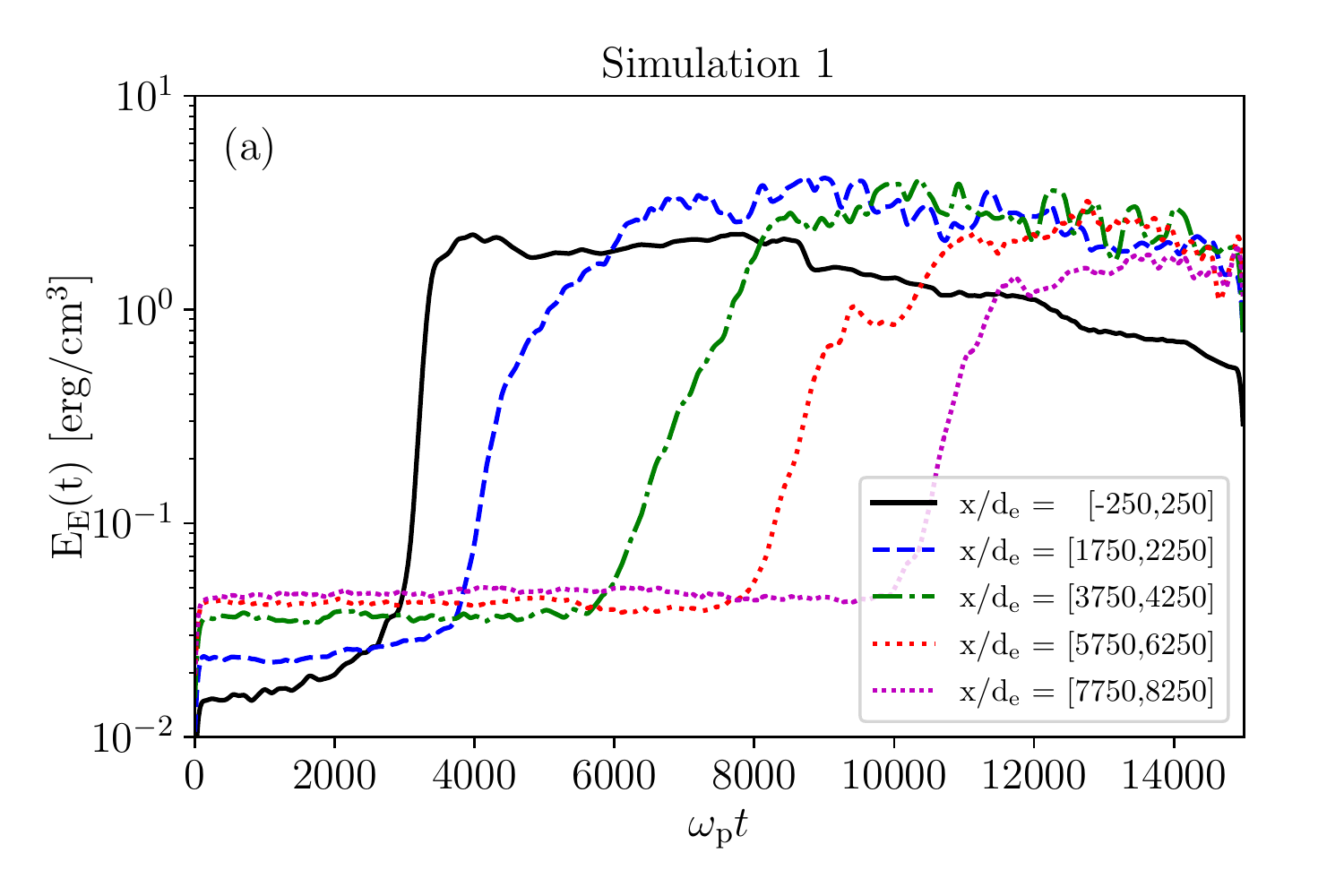}
    \includegraphics[width=0.49\textwidth]{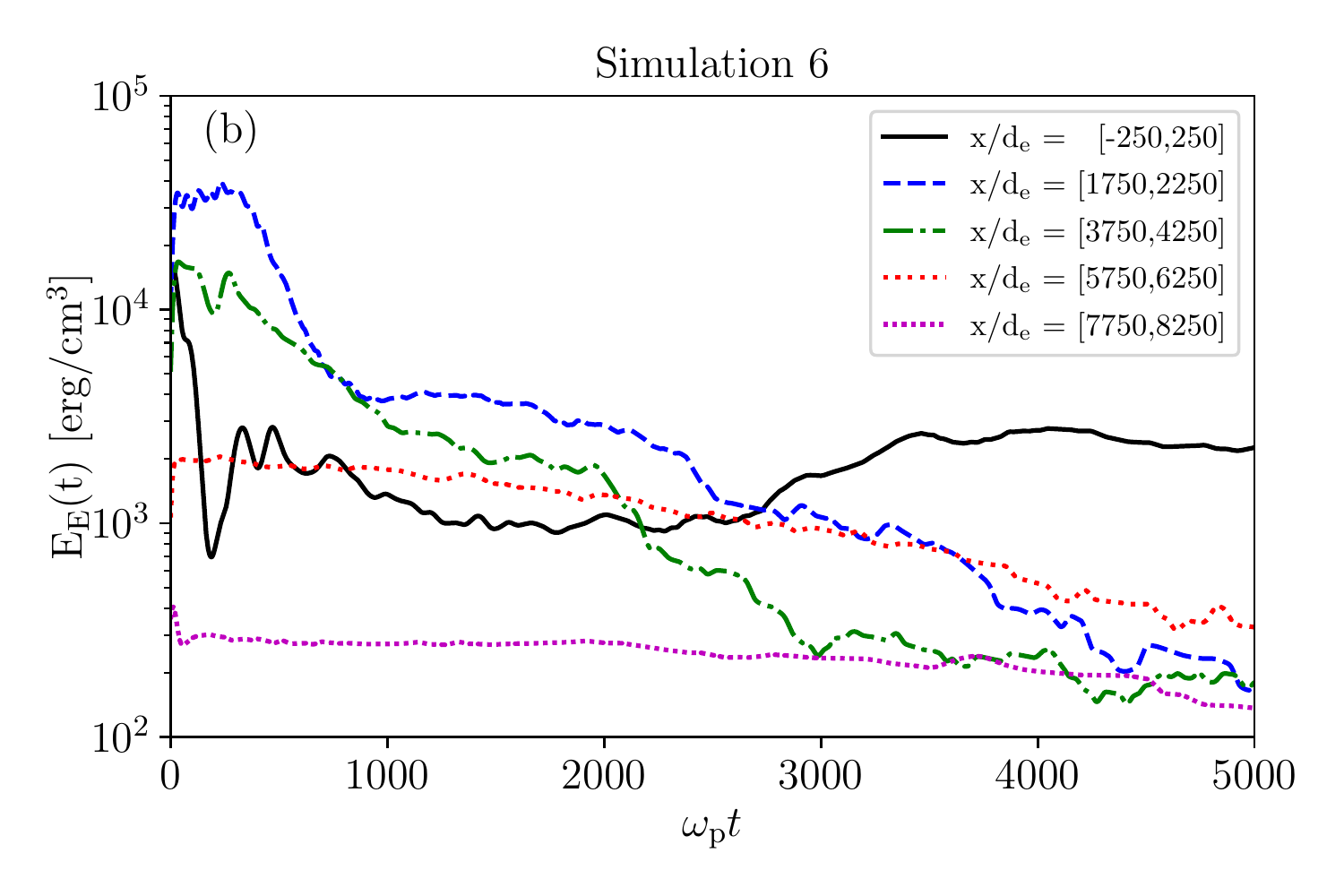}
    \caption{Evolution of the electrostatic energy density in selected regions with increasing distance from the simulation center
    in Simulation~1 (a) and Simulation~6 (b).
    See text for growth rates in Simulation~1.}
    \label{fig:energy-along-x}
\end{figure*}

The evolution of the mean electrostatic energy density in selected regions along $x$ is shown in Figure~\ref{fig:energy-along-x}.
In Simulation~1, the initial energy density evolution is not the same for all positions.
It increases from the simulation center to the edge as the particle density increases.
As the instability starts at the center, the energy density starts to growth there first.
Then, further regions manifest exponential energy increase as they are further away from the center.
The estimated growth rates $\Gamma/\omega_\mathrm{p}$ of the exponentially growing parts are:
$(4.43 \pm 0.05) \times 10^{-3}$ for $x/d_\mathrm{e} = [-250,250]$;
$(2.28 \pm 0.03) \times 10^{-3}$ for $x/d_\mathrm{e} = [1750,2250]$;
$(1.24 \pm 0.02) \times 10^{-3}$ for $x/d_\mathrm{e} = [3750,4250]$;
$(1.61 \pm 0.03) \times 10^{-3}$ for $x/d_\mathrm{e} = [5750,6250]$;
and $(1.37 \pm 0.02) \times 10^{-3}$ for $x/d_\mathrm{e} = [7750,8250]$.
The growth rate decreases along $x$ because the particle velocity distributions
broadens, and the two-stream instability weakens.
However, the saturation energy density is almost the same in all regions,
reaching the mean energy density of $\sim 2$~erg/cm$^{3}$.
After the start of Simulation~6, most of the electrostatic energy density is generated 
in $x/d_\mathrm{e} = [1750,2250]$,
where their mean energy density reaches $\sim 3.5\times 10^4$~erg/cm$^{3}$.
Although the energy density is not constant along $x$ in the selected boxes,
the mean value along a sufficiently large region can be a good representative.
The time evolution of the energy densities has a decreasing trend with the exception
of the simulation center $x/d_\mathrm{e} = [-250,250]$, where the energy density varies in time and remains $> 10^3$~erg/cm$^{3}$ until the simulation end.

\subsubsection{Wave Profiles}
\begin{figure*}[!ht]
    \centering
    \includegraphics[width=0.49\textwidth]{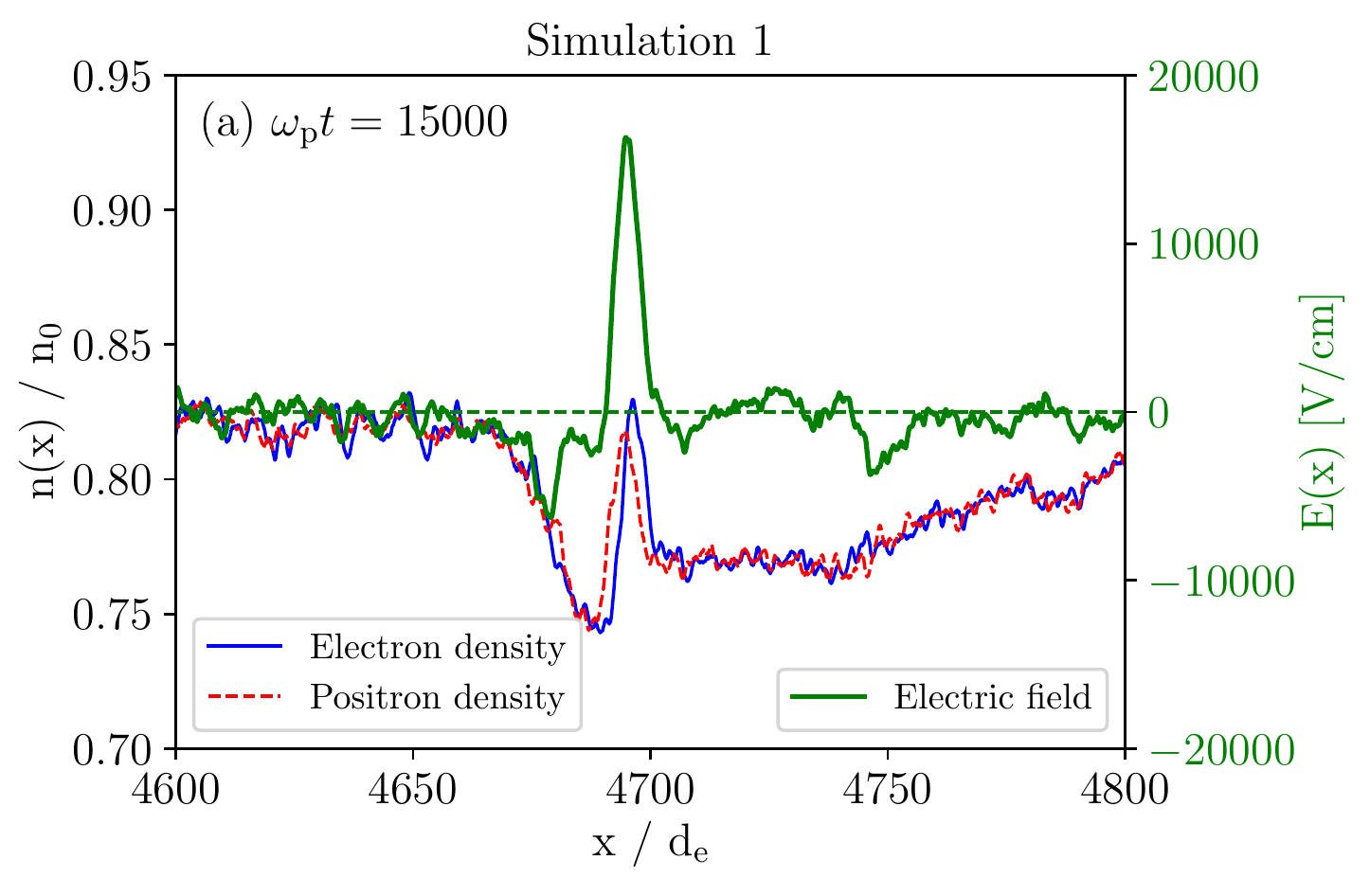}
    \includegraphics[width=0.49\textwidth]{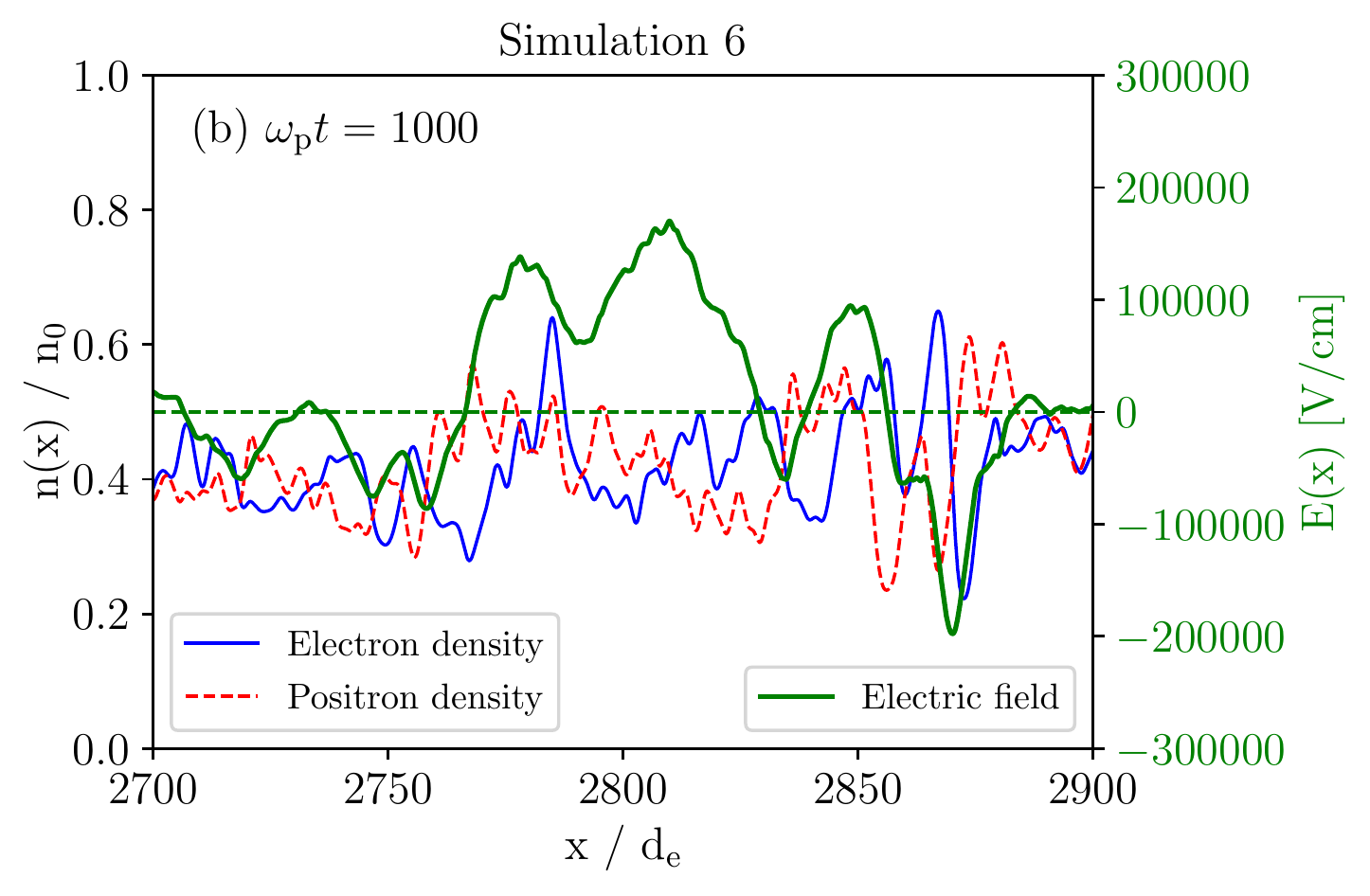}
    \caption{Examples of profiles of the electron and positron particle density,
    and electrostatic field component as functions of distance $x$ from the simulation center.
    (a) Simulation~1 at $\omega_\mathrm{p}t = 15000$.
    (b) Simulation~6 at $\omega_\mathrm{p}t = 1000$.}
    \label{fig:field-along-x}
\end{figure*}

Figure~\ref{fig:field-along-x} shows two example profiles of the
electron and positron density and the intensity of the electric field.
We select typical regions where one of the largest electrostatic waves appears.
For Simulation~1, the wave electric intensity reaches $\sim 15$~kV/cm.
At the same position, the electrons and positrons form a density fluctuation.
The particle densities to the left from the wave are almost constant,
only small fluctuations are visible.
On the right from the wave center, the particle density decreases with respect to the left side.
In Simulation~6, the density and the electric intensity vary more strongly along $x$.
The maximal intensity of the electric field reaches $\sim -200$~kV/cm.
The density varies in the range $0.3-0.65\,n_0$
and they are out of phase between electrons and positrons.

\subsubsection{Particle Phase Space for Selected Positions}
\begin{figure*}[!ht]
    \centering
    \includegraphics[width=0.49\textwidth]{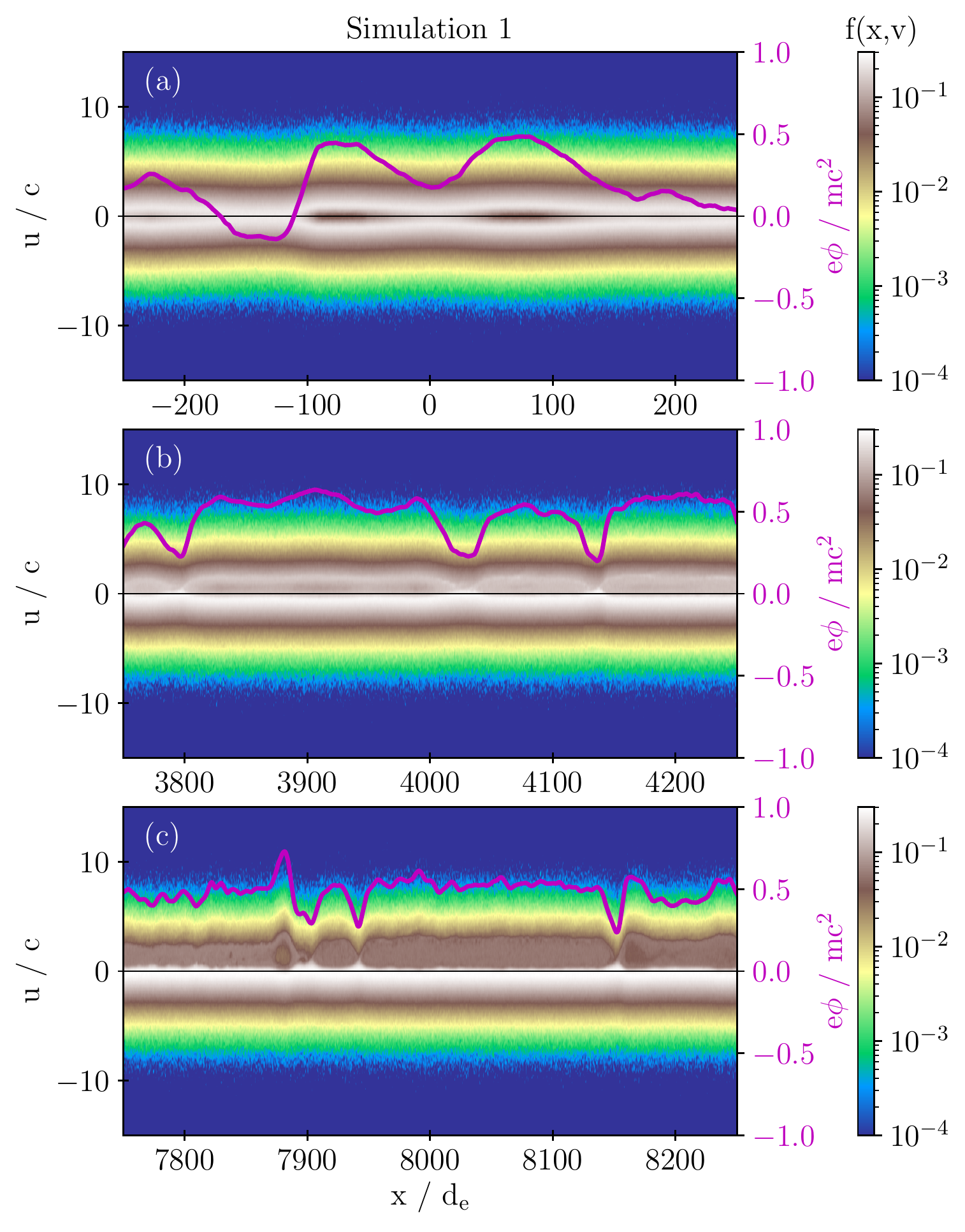}
    \includegraphics[width=0.49\textwidth]{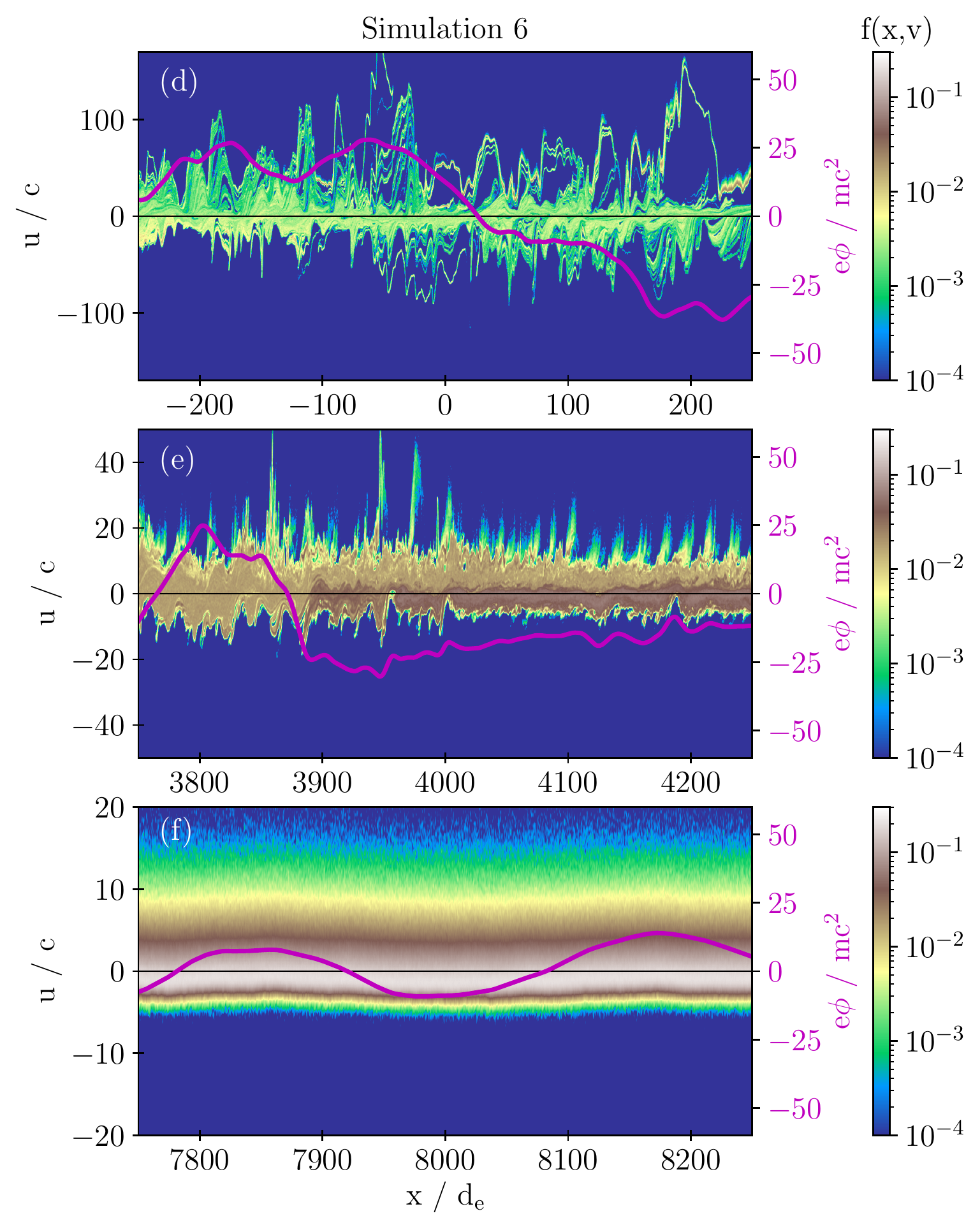}
    \vspace{-0.3cm}
    \caption{Electron phase spaces at selected distances from the simulation center
    overlaid by the electric potential
    (magenta line).
    (a) Simulation~1 at $\omega_\mathrm{p}t = 15000$.
    (b) Simulation~6 $\omega_\mathrm{p}t = 1000$.
    See Figure~\ref{fig:distr-detail} for velocity profiles.}
    \label{fig:phasespace-detail}
\end{figure*}

Figure~\ref{fig:phasespace-detail} shows details of the electron phase spaces in selected regions along the $x$-direction.
In Simulation~1 and $x/d_\mathrm{e} = [-250,250]$ (Figure~\ref{fig:phasespace-detail}a), the distribution is symmetric with respect to the axis $u/c = 0$. 
Most of the particles with $u/c < 0$ initially come from the right bunch,
while those with $u/c > 0$ from the left bunch.
The electric potential energy amplitude is $\sim 0.35\,m_\mathrm{e}c^2$.
In $x/d_\mathrm{e} = [7750,8250]$ (Figure~\ref{fig:phasespace-detail}c), particles with $u/c \leq 0$ dominates.
The electric potential energy amplitude is $\sim 0.18\,m_\mathrm{e}c^2$.
In Simulation~6, the plasma is heated in $x/d_\mathrm{e} = [-250,250]$.
The electric potential energy varies with amplitude $\sim 50\,m_\mathrm{e}c^2$.
In $x/d_\mathrm{e} = [3750,4250]$, most of the particles are localized 
in $u/c = [-15,15]$.
At the position $x/d_\mathrm{e} \sim [3850,3880]$, a strong potential wall is formed
with an electric potential energy height of $\sim 32\,m_\mathrm{e}c^2$.
In $x/d_\mathrm{e} = [7750,8250]$, the distribution is smooth,
no phase space holes are formed, and the electric potential energy varies
 with amplitude $\sim  13\,m_\mathrm{e}c^2$.

\subsubsection{Velocity Distributions}
\begin{figure*}[!ht]
    \centering
    \includegraphics[width=0.49\textwidth]{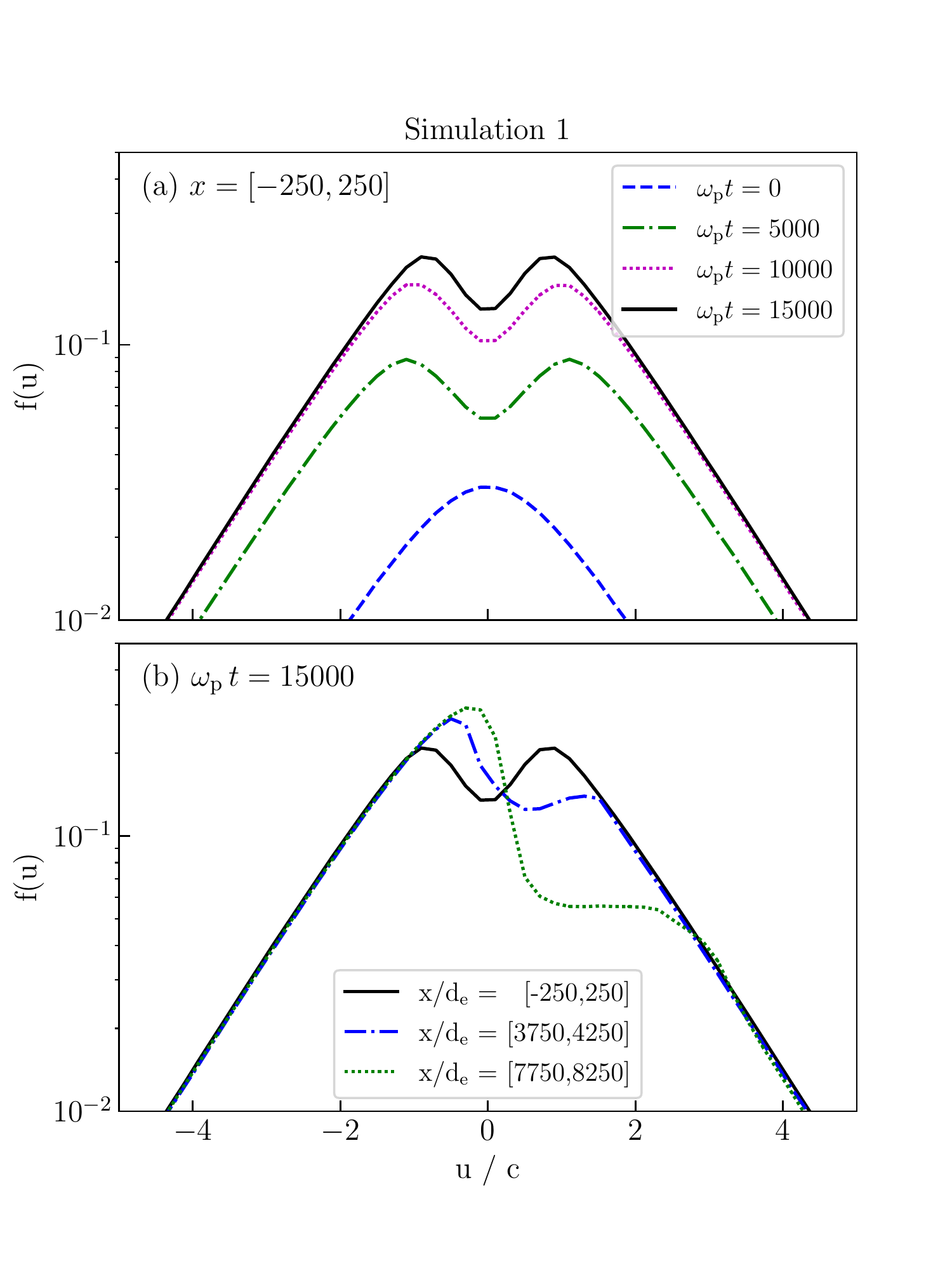}
    \includegraphics[width=0.49\textwidth]{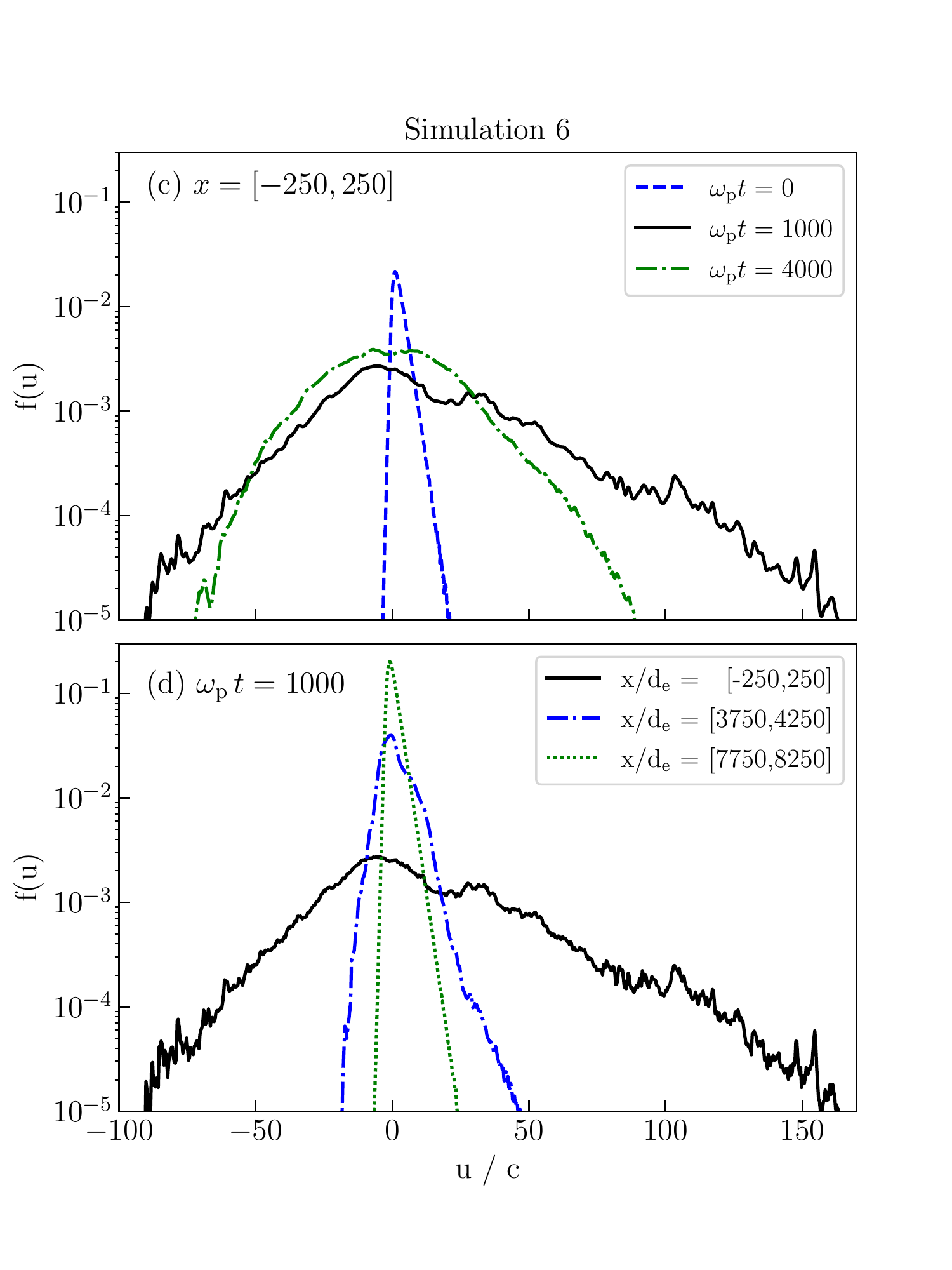}
    \caption{
    Electron velocity distributions as a function of time and distance from the simulation center:
    (a-b)~Simulation~1, (c-d)~Simulation~6.
    (a,c) $x/d_\mathrm{e} = [-250,250]$ at selected time moments.
    (b) Selected regions at $\omega_\mathrm{p} t = 15000$.
    (d) Selected regions at $\omega_\mathrm{p} t = 1000$.
    }
    \label{fig:distr-detail}
\end{figure*}

The mean velocity distribution functions $f(u,x)$ in selected time steps and regions are shown for electrons in Figure~\ref{fig:distr-detail}.
The velocity $u$ is the only non-zero component of the particle velocity vector $\vec{u} = (u,0,0)$.
Distributions are normalized as $\int_{-\infty}^{\infty} f(u,x) \mathrm{d}u = n(x) / n_0$.
In Simulation~1 and $x/d_\mathrm{e} = [-250,250]$ (Figure~\ref{fig:distr-detail}a), the plasma is heated and the electron number density is enhanced.
The distributions show a double-peaked profile from the start of the particle
overlap until the simulation end.
The distribution function profile also changes along the simulation (Figure~\ref{fig:distr-detail}b).
In the simulation center, both distribution peaks have the same value.
The same amount of electrons comes from the left and right bunch into the simulation center.
At $x/d_\mathrm{e} = [3750,4250]$,
more electrons from the right bunch are present and both distribution maxima shift towards higher positive velocities.
At a distance $x/d_\mathrm{e} = [7750,8250]$ from the center,
only one peak at negative velocities and plateau for positive velocities are present.
On the other hand, at the center of Simulation~6, the plasma is quickly heated after the simulation start.
Electrons also arrive into the simulation center during the evolution.
Though we select exactly the simulation center,
at $\omega_\mathrm{p}t = 1000$ the tail of the distribution contains significantly more electrons with positive velocities.
Despite the limited number of macro-particles in the simulation,
the distribution tail is covered well. 
Moreover, $\sim 0.1\,\%$ of electrons exceed the velocity $u/c = 150$.
At later simulation times, the plasma is slightly cooled down, and the distribution becomes more thermalized with a mean value $\sim 0 c$, but still significantly hotter than the initial plasma.
Moreover, the distribution function also significantly varies with the distance from the simulation center.
While the plasma is hot in the simulation center,
it becomes cooler with increasing distance from the simulation center.

\subsubsection{Electrostatic Wave Dispersion along Simulation Domain}
\begin{figure*}[!ht]
    \centering
    \includegraphics[width=0.49\textwidth]{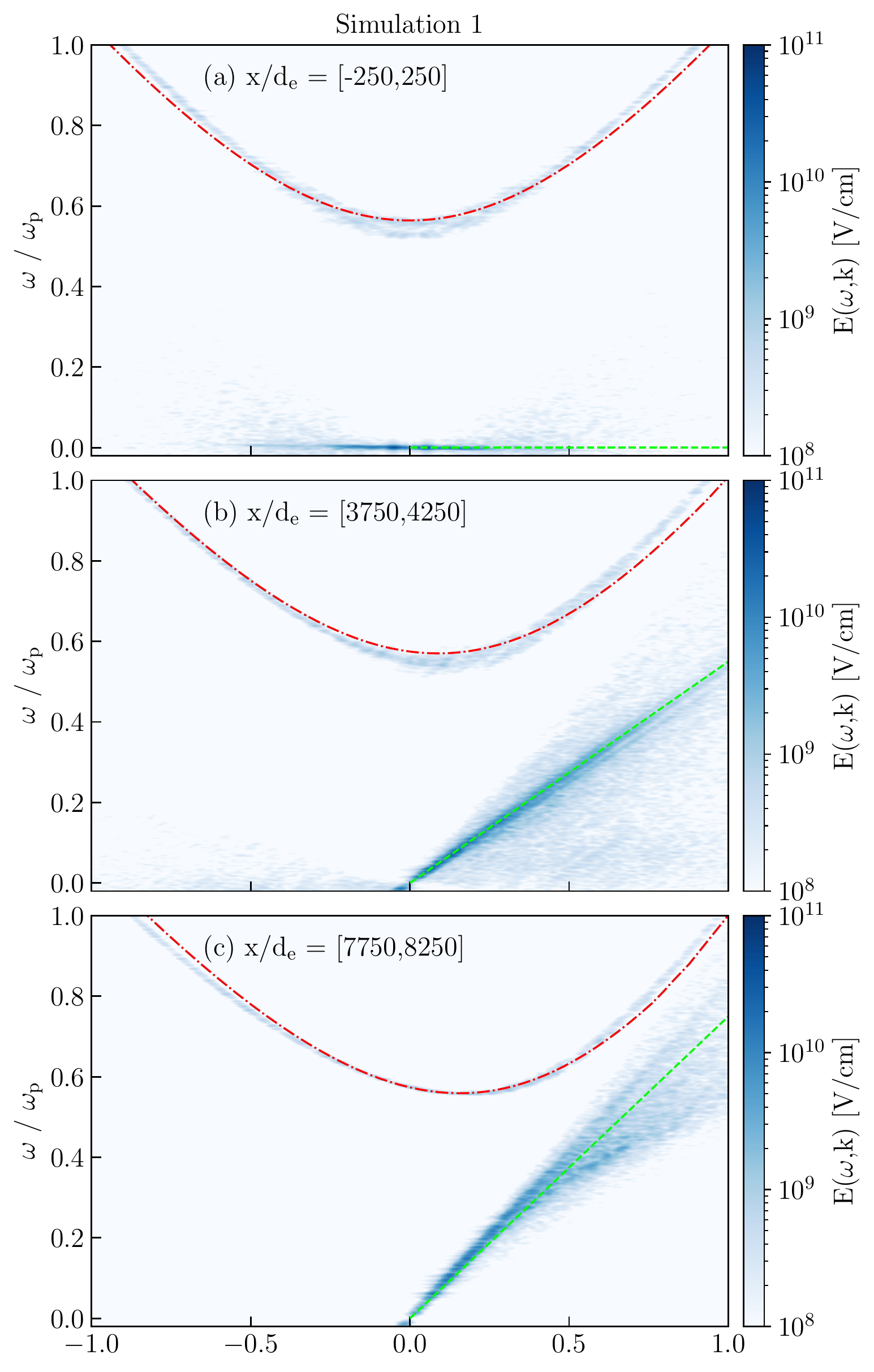}
    \includegraphics[width=0.49\textwidth]{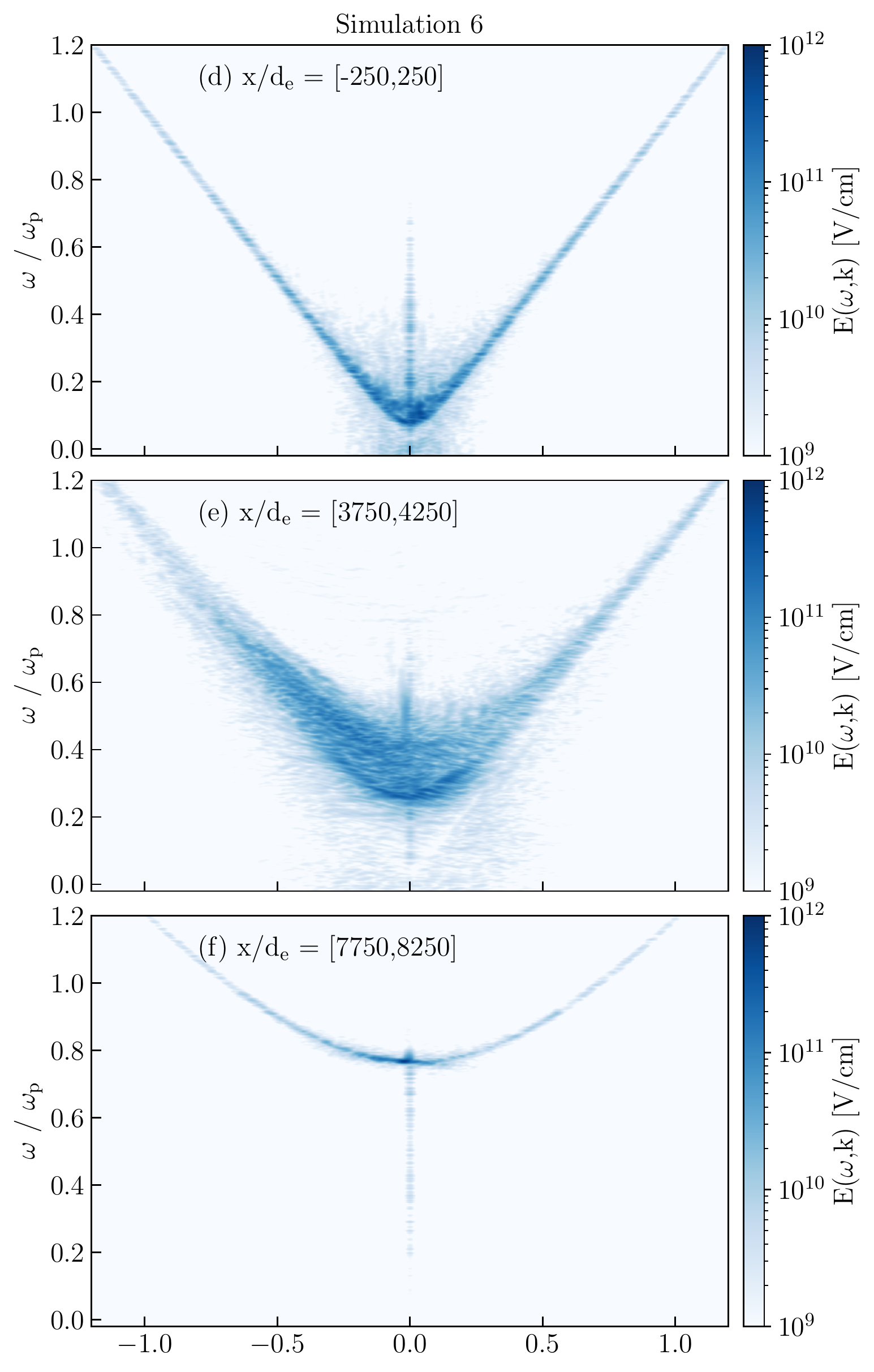}
    \caption{
    Dispersion diagrams for selected regions.
    (a) Simulation~1 at $\omega_\mathrm{p} t = 12500-15000$.
    Dispersion diagrams are overlaid by relativistically shifted
    superluminal L-mode waves (red dashed line) for phase speeds
    $v/c = 0, 0.3, 0.5$ (from tom to bottom)
    and subluminal beam-mode waves (green dashed line) for
    $v/c = 0, 0.55, 0.75$~$c$ assuming a Maxwell-J\"uttner distribution.
    (b) Simulation~6 at $\omega_\mathrm{p} t = 0-2500$.
    }
    \label{fig:phasespace-detail-sim1}
\end{figure*}

Figure~\ref{fig:phasespace-detail-sim1} shows dispersion diagrams for selected regions along the $x$ axis.
As these regions are smaller than in Figures~\ref{fig:fourierevolution1}-\ref{fig:fourierevolution3},
they do not cover such a high range of plasma densities,
and the broadening of the superluminal L-mode is smaller.
In Simulation~1 for $x/d_\mathrm{e} = [-250,250]$, the enhanced subluminal L-mode waves  have group velocity $u/c \sim 0$.
It velocity increases with increasing distance from the simulation center and approaches the speed of light.
Also the superluminal L-mode waves show up the relativistic shift
as the plasma drifts towards the simulation center with relativistic velocities.
Figures (a-c) are overlaid by analytical dispersion solutions of superluminal and subluminal L-mode waves for Maxwell-J\"{u}ttner distribution with $\rho = 1$.
Relativistic corrections with $v/c = 0, 0.3, 0.5$ are applied to superluminal L-mode waves.
The subluminal waves are overlaid by beam-mode waves with relativistic corrections $u/c = 0,0.55,0.75$.
In Simulation~6, the frequency of the superluminal L-mode waves increases with increasing distance from the simulation center.

%%%%%%%%%%%%%%%%%%%%%%%%%%%%%%%%%%%%%%%%%%%%%%%%%%%%%%%%%%
%%%%%%%%%%%%%%%%%%%%%%%%%%%%%%%%%%%%%%%%%%%%%%%%%%%%%%%%%%
\section{Discussion} \label{sec:disc}
A number of proposed pulsar radio emission models are based on the
consideration of plasma instabilities due of interacting streaming
bunches of positrons and electrons in the lower parts of
pulsar magnetospheres.
In order to verify this conjecture we conducted a series of PIC-code 
simulations to comprehend the nonlinear evolution of the instabilities
of a corresponding relativistically hot pair plasma for parameters 
appropriate to describe this situation: for a range of
initial particle thermal velocities, of distances between the bunches, 
and the drift velocities between electrons and positrons.
We found that the initial drift velocity between the particle species 
is the main parameter governing the bunch evolution.

If there is no initial relative drift, the bunches just expand until
the fastest particles of two bunches reach each other to cause,
eventually, a streaming instability.

If there is a finite initial drift between the species, however, 
the expansion becomes suppressed since in this case strong 
electrostatic fields are generated. As a result the plasma becomes 
strongly heated, the particles become confined and no
streaming instability takes place when the bunches overlap.

Already previous studies revealed an important constraint 
for the bunch interaction mechanism to work:  
a sufficiently long time interval is needed to allow bunch 
expansion and interaction. This condition was found not to be 
fulfilled in typical pulsar magnetospheres~\citep{Melrose2020a}.

We now found an additional necessary condition constraining
the bunch interaction mechanism: 
the velocity space distributions of both electrons and positrons
have to be similar all along the whole bunch. Equivalently there should be
no mutual drift of the bunches and the density profiles along 
the bunch must be similar.
These conditions very much constrain the opportunities
of the interacting bunch mechanism to work since if they
are not fulfilled immediately strong electrostatic waves are 
generated which prevent any bunch expansion.
As the pulsar magnetosphere is a very dynamic and the pair 
bunches are created in a cascading process it is statistically 
very improbable that particles are distributed in a way 
needed for the instability to develop.

For a finite mutual drift speed of the electrons and positrons 
the bunch would expand at different velocities. Thus they 
cause the formation of local discharges associated with an 
ambipolar electric field at density gradients at
timescales $\sim \omega_\mathrm{p}^{-1}$.
As both particle species have relativistic velocities,
strong ambipolar electric fields are generated
which prevent any particle expansion while
locally heating the plasma.
Even for a small initial drift velocity $u_\mathrm{d}/c = 0.1 < 1/\sqrt{\rho}$,
the amount of released electrostatic energy increases by three orders of magnitude 
compared to the case of vanishing initial drifts.
In dependence on the initial drift velocity,
the energy density of waves in Simulations~5-7 reached up to  
$E_\mathrm{E} \sim (0.017-2.5) \times 10^{5}$~erg/cm$^{3}$ 
($E_\mathrm{E} / (n_0 m_\mathrm{e} c^2) \sim (0.066 - 9.6) \times 10^{-1}$)
and the maximal electric field amplitude up to 
$E \sim (0.62-7.5) \times 10^{5}$~V/cm 
($E / (m_\mathrm{e} c \omega_\mathrm{p} e^{-1}) \sim (0.37 - 4.4)$).

At the bunch center, however, a streaming instability between electrons and 
positrons can take place.
The amount of electrostatic energy built up by this instability is,
however, much smaller than the amount of energy generated 
due to the bunch gradient.

In the case of a vanishing initial drift and if both species 
are initially equally distributed in the velocity space and 
at all locations, both species would expand symmetrically,
the fastest particles (in the bunch reference frame) would
escape the bunch.
They can interact with the background plasma,
forming a weak streaming instability as they propagate.
Later they ``catch up'' with fast particles of the adjacent 
bunch (expanding in the opposite direction in the bunch reference frame).
They cause a two-stream instability that saturates 
at $\sim 1000\,\omega_\mathrm{p}t$ after their overlap
and eventually form electrostatic waves that survive until the simulation end.
In dependence on the initial plasma temperature,
the electrostatic wave energy density is in range 
$E_\mathrm{E} \sim (0.14-1.6) \times 10^{2}$~erg/cm$^{3}$
($E_\mathrm{E} / (n_0 m_\mathrm{e} c^2) \sim (0.53 - 6.1) \times 10^{-4}$)
and electric field amplitude 
$E \sim (0.56-1.9) \times 10^{4}$~V/cm
($E / (m_\mathrm{e} c \omega_\mathrm{p} e^{-1}) \sim (0.33 - 1.1) \times 10^{-1}$).

The growth rate values of the instability in the simulation center 
are similar to those found by \citet{Manthei2021}, despite 
their approximation of the bunch interaction by beam and background 
plasmas described by a relativistic Maxwell-J\"{u}ttner velocity space
distribution. 
Moreover, we also found that the growth rate is decreasing with an 
increasing distance from the simulation center due a change 
of the shape of the velocity distribution function from an unstable 
double-peaked to a stable peak-plateau distribution.

Taking all effects of the bunch expansion into account, we found that
electrostatic waves are formed, independent
on the initial plasma temperature
if the particle overlap in phase space and streaming instability are allowed.
They are formed even in a highly relativistic plasma with $\rho \ll 1$.
Note that this finding is different from the temperature range $\rho \geq 1.66$ of wave formation in a uniform streaming instability found by \citet{Benacek2021}.
This also answers the question raised in a review of pulsar emission 
processes by\citet{Melrose2020b}: whether the velocity space distributions 
of overlapping bunch particles are sufficiently well separated to become 
unstable to generate waves.

We also found that the most intense electrostatic waves 
differ between the cases without and with finite initial drift velocity.
For a zero initial drift velocity, the subluminal L-mode waves 
(relativistic generalization of the Langmuir waves) with phase velocity $v_\phi = 0 - 0.75\,c$ are excited.
Superluminal waves are faint in a low temperature plasma;
however, their intensity increases with increasing plasma temperature.
For a nonzero drift velocity, the superluminal L-mode waves are the most intense waves and the subluminal modes are weak.

For the waves we obtained by our simulations 
(see, e.g.,  the dispersion diagrams in Figures~\ref{fig:fourierevolution1}-\ref{fig:fourierevolution3}), 
several electromagnetic emission mechanisms could 
apply:

First electromagnetic waves could be directly generated by a linear acceleration 
mechanism.
This mechanism would be more efficient in the case of finite 
initial drifts between species, in which strong electrostatic fields are 
generated~\citet{Melrose2017a,Melrose2020a}.
An oscillation frequency close to the plasma frequency 
was suggested to be the emission frequency~\citep{Eilek2016} .
We found, instead, that over a broad frequency range
waves are generated.
The local superluminal L-mode frequency decreases as $\langle \gamma^{-3} \rangle$ with temperature.
The typical found frequencies are in the range $\omega/\omega_\mathrm{p} \sim [0.01,0.5]$,
i.e., a factor $2-100$ times smaller than the plasma frequency.
The motion of individual particles in such a field must be taken into an account to 
properly quantify this emission process. 
The prediction of emission properties is non-trivial as there can be several 
types of emitting particles, e.g., particles that can be captured in electric 
fields and particles with a kinetic energy large enough to pass through 
such the electrostatic waves.
The situation is more complicated due to the fact that the oscillation frequency can be  smaller than the local plasma frequency, but the surroundings of the bunch can be sufficiently dilute 
to allow the propagation of electromagnetic waves.
The relativistic frequency shift further increases the emission frequency.

A second plausible mechanism is a relativistic plasma emission.
In this process L-mode waves interact with other waves, e.g.,
other L-mode waves or density waves, to generate electromagnetic 
waves.
One option is that the subluminal L-mode waves propagate 
in a given direction interact with a counter-propagating wave.
However, in all our simulations, no sufficiently intense waves 
were generated propagating in the opposite direction at
a sufficiently high frequency and intensity. Instead, the 
subluminal L-mode waves mostly propagate along the same 
direction. 
Another possibility would be a modulation by an external 
wave with a component oscillating perpendicular to the 
magnetic field~\citep{Weatherall1997}. 
Such electromagnetic waves could escape the emission region.

Other plasma emission mechanisms --- the free electron maser, 
curvature emission, electron-cyclotron maser --- do not apply 
to instabilities caused by a bunch-bunch interaction.
 Curvature emission assume independent beams, not necessarily an interaction among them.
 Free electron maser- and electron-cyclotron maser-mechanisms 
 require a finite energy of perpendicular particle motion
 and specific types of velocity space distribution functions
 which cannot be assumed to be formed in the lower pulsar 
 magnetosphere.

%%%%%%%%%%%%%%%%%%%%%%%%%%%%%%%%%%%%%%%%%%%%%%%%%%%%%%%%%%%%%%%%%%%%%%%%%%%%%%%%%%%%%%%%%%%%%%%%%%%%%%%%%%%%%%
\section{Conclusions} \label{sec:conc}

We studied the nonlinear evolution and interaction of electron-positron bunches 
(clouds) in pulsar magnetospheres in dependence on the relative drift speed
of electrons and positrons, the plasma temperature, and the distance between 
the bunches.
The resulting plasma instabilities and particle oscillations are hoped to
cause coherent radio emissions.

We found that already a small drift speed between electrons and positrons
causes the generation of large amplitude electrostatic superluminal L-mode 
waves.
The wave energy reaches up $~15\,\%$ of the total kinetic particle energy.
These waves heat the plasma, but no streaming instability between the bunches is triggered.
For a drift velocity $u_\mathrm{d}/c = 10$, the electrostatic energy density can 
locally reach up to $E_\mathrm{E} \sim 2.5 \times 10^{5}$~erg/cm$^{-3}$ 
($E_\mathrm{E} / (n_0 m_\mathrm{e} c^2) \sim 0.96$)
and a electric field intensity can become as large as $E \sim 7.5 \times 10^{5}$~V/cm
($E / (m_\mathrm{e} c \omega_\mathrm{p} e^{-1}) \sim 4.4$).
For zero drift speeds between the bunches they overlap in phase space and 
cause streaming instabilities.
The generated electrostatic waves are weaker compared to the precedent scenario.
Their electrostatic energy density reach up to $E_\mathrm{E} \sim 1.6 \times 10^{2}$~erg/cm$^{3}$
($E_\mathrm{E} / (n_0 m_\mathrm{e} c^2) \sim 6.1 \times 10^{-4}$)
and electric field intensity $E \sim 1.9 \times 10^{4}$~V/cm
($E / (m_\mathrm{e} c \omega_\mathrm{p} e^{-1}) \sim 1.1 \times 10^{-1}$).
Such regime very much requires, however, the same velocity space distribution 
of both species all along the whole bunches - which further constrains the 
applicability conditions of this mechanism.

\acknowledgments


The authors acknowledge the support by the German Science Foundation (DFG) projects BU 777-17-1 and MU-4255/1-1. 
We acknowledge the developers of the ACRONYM code (Verein zur F\"orderung kinetischer Plasmasimulationen e.V.).
The authors gratefully acknowledge the Gauss Centre for Supercomputing e.V. (\url{www.gauss-centre.eu}) for partially funding this project by providing computing time on the GCS
Supercomputer SuperMUC-NG at Leibniz Supercomputing Centre (www.lrz.de), projects pr74vi and pn73ne.
This work was supported by the Ministry of Education, Youth and Sports of the Czech Republic through the e-INFRA CZ (ID:90140).
Part of the simulations was carried out on the HPC-Cluster of the
Institute for Mathematics of the TU Berlin.
The authors also thankfully acknowledge their enlightening discussions with Axel Jessner of
the Max-Planck-Institute for Radioastronomy in Bonn.

\software{PIC-code ACRONYM, Python}



% \appendix
%\section{Appendix information}
% The time needed for catching up slow particles of the leading bunch is 
% sufficiently small, but the time for catching up with particles moving 
% at the mean velocity is too long.


\bibliography{references}{}
\bibliographystyle{aasjournal}



\end{document}